# Noise Model of Relaxation Oscillators Due to Feedback Regeneration Based on Physical Phase Change

Bosco Leung
Electrical and Computer Engineering Department, University of Waterloo
Waterloo, Ontario, Canada
bleung@uwaterloo.ca

*Abstract*— **A new approach to investigate noise spikes due to regeneration in a relaxation oscillator is proposed. Noise spikes have not been satisfactorily accounted for in traditional phase noise models. This paper attempts to explain noise spikes/jump phenomenon by viewing it as phase change in the thermodynamic system(for example, from gas to liquid or magnetization of ferromagnet). Both are due to regeneration (positive feedback in oscillator as well as alignment of spin due to positive feedback in ferromagnet). The mathematical tool used is the partition function in thermodynamics, and the results mapped between thermodynamic system and relaxation oscillator. Theory is developed and formula derived to predict the magnitude of the jump, as a function of design parameter such as regeneration parameter or loop gain. Formulas show that noise increases sharply as regeneration parameter/loop gain approaches one, in much the same way when temperature approaches critical temperature in phase change. Simulations on circuits (Eldo) using CMOS as well as Monte Carlo simulations (Metropolis) on ferromagnet (Ising model) were performed and both show jump behaviour consistent with formula. Measurements on relaxation oscillators fabricated in 0.13um CMOS technology verify such behaviour, where the sharp increase in noise when regeneration parameter/loop gain is close to one, matches closely with the theoretical formula. Using the formula the designer can quantify the variation of noise spikes dependency on design parameters such as $g_m$ (device transconductance), R, $I_0$, via their influence on regeneration parameter/loop gain.**

## I. INTRODUCTION

Relaxation oscillators are used in applications [1], [2] such as phase locked loops for clock recovery, clock generators, and time-to-digital converters. Unlike LC oscillators, these oscillators do not need inductors and are easily implemented in CMOS process. While simple in implementation, their phase noise have been inferior. Traditional circuit phase noise models based on linear oscillatory systems [3], impulse sensitivity function [4], as well as those based on threshold crossing have been applied to relaxation and ring oscillators [5], [7], [10].

These approaches are not suitable because, as result of nonlinear regeneration, relaxation oscillator's noise trajectories can create noise spikes [8], particularly at time of switching, the moment at which timing jitter is to be calculated. Model in [8], [9], try to address this issue, using the stochastic theory of large deviation [19] to find the bound of the noise spikes. For each sample path of the noisy trajectory, such a bound depends on $I_\varepsilon$, the action function associated with that trajectory. However near the jump point, [8] admits that current *i* can have deviation approaching infinity (after eq. 5.7 in [8]). Therefore [8] has severe drawback (see Remarks (3), pg. 662 [8]). This is because $I_\varepsilon$ is an integral criterion and the noisy trajectory is a badly behaved set(remark 2 after eq. 5.5 in [8]).

[21] went back to [9], from which [8] is based. [9] made a first attempt of explaining the jump phenomenon as phase change (like, from gas to liquid). [21] now investigated the phase change based jump phenomenon in relaxation oscillator using such physics concept as the Lagrangian As pointed out by [23], [21] is missing a dissipation term, which is needed to account for effect from, for example, the dissipation effect from resistor. [22] incorporates such effect, using the Rayleigh dissipation function [24] and Lagrangian with friction [25]. As pointed out by [25], this dissipation based phase change formulation, is fundamentally different from former model of nonlinear circuit [26] (Taylor approach of Stratonovich), [27]. Following [5], this report, as well as [21], [22], focus on thermal noise and MOS circuits.

Due to length restriction only preliminary and qualitative aspect was discussed in [21], [22]. Also the variational approach developed in [21], [22] cannot be directly applied. The present report fully developed [21], [22] and overcome such limitation. It now gives a full model, with quantitative closed form formula and experimental results. To find fluctuation, instead of applying variational approach through the entire trajectory, as in [8], [21], [22], we now focus around the point in the trajectory where jump point/phase change occurs. This can be done, in the vdw gas, by using the partition function. The fluctuation calculated from the partition function is then mapped back to relaxation oscillator. The tool used in mapping between the relaxation oscillator and vdw gas is by comparing the action based on Lagrangian formulation between the relaxation oscillator action $I_\varepsilon$, and the vdw gas action based on the corresponding Lagrangian. The way of obtaining fluctuation formula is from recognizing that interaction (due to poisitive feedback) is responsible for overcoming the noise to achieve regeneration, thus resulting in a low entropy state (capacitor charged or discharged).

Section II reviews the previous model. Section III starts to develop the noise model, by focusing on the qualitative aspect. Equations describing the circuits and vdw gas are set up and qualitative behaviour on the probability distribution of resulting noise/fluctuation is presented. Section IV derives the noise/fluctuation formula. It first derives the noise/fluctuation formula of vdw gas, based on the state equation describing the vdw gas set up in Section III, and the partition function. It then maps this noise/fluctuation formula of vdw gas into the relaxation oscillator. Design example is presented. Section V then presents the simulation results and experimental results (chip fabricated in 0.13um CMOS), which verify the theory. Conclusions are drawn in section VI.

## II. RELAXATION OSCILLATOR: PREVIOUS MODELS

The source coupled multivibrator is one of the most popular relaxation oscillator and we choose it for our investigation [2], [7]. Shown in Figure 1 is such a realization. (Another one is theical capacitor configuration (fig. 16 of [7])). However regeneration behaviour is similar in both, and both have noise spikes [1]. Thus we concentrate on Figure 1 as a representative example. First neglecting noise source $i_n$, the circuit equations are:

$$\frac{dV}{dt} = \frac{I_0 - i}{C} \tag{1}$$

$$V - \left(\sqrt{\frac{2i}{k_n(W/L)}} - \sqrt{\frac{4I_0 - 2i}{k_n(W/L)}}\right) - (2I_o - 2i)R = 0 \tag{2}$$

Continuous solution exists so long as current $I$ can be solved continuously as a function of $V$, so as to obtain:

$$\frac{di}{dt} = \frac{\frac{I_0 - i}{C}}{\left(\left(\frac{1}{2}\left(\frac{2i}{k_n(W/L)}\right)^{-\frac{1}{2}} + \frac{1}{2}\left(\frac{4I_0 - 2i}{k_n(W/L)}\right)^{-\frac{1}{2}}\right)\frac{2}{k_n(W/L)} + 2R\right)} \tag{3}$$

At some point the denominator becomes zero and the solution is ill posed. Following [8], regularization is then accomplished by taking into account the fact that parasitic capacitances present in the transistors. Then (2) becomes:

$$\varepsilon \frac{di}{dt} = V - \left(\sqrt{\frac{2i}{k_n(W/L)}} - \sqrt{\frac{4I_0 - 2i}{k_n(W/L)}}\right) - (2I_o - 2i)R \tag{4}$$

Here $\varepsilon$ is a small parameter that accounts for parasitics in the circuit. Next we Taylor expand (4), keep the linear, cubic terms. We then normalize time $t$ by $RC$, so that in (1), (4) the variable $V$ is written as $x$, and variable $i$ is written as $y$. This also means the coefficients of the linear and cubic terms are dimensionless. For ease of illustration, for now we select $I_0$, $R$, $(W/L)_{1-2}$ so that the coefficients have values of one. The result is:

$$\begin{aligned}\dot{x} &= f(x, y) \\ \varepsilon \dot{y} &= g(x, y)\end{aligned} \tag{5}$$

where $f(x,y)=y$, $g(x,y) = -x-y^3+y$ 

$$\tag{6}$$

Next we add noise to the circuit. We assume all the noise is lumped, as shown in 'noise i' in Figure 1. Following notation in [8] (eq. 4.11), we further assume the noise term is written as $(2\sqrt{\lambda})i_n$, where $\lambda$ is the intensity of the noise source, with $i_n$ having unit strength (e.g for a resistor R, $\lambda$ is kT/R, $2\sqrt{\lambda}$ is $\sqrt{(4kT/R)}$). Then (2) becomes :

$$V - \left(\sqrt{\frac{2i}{k_n(W/L)}} - \sqrt{\frac{4I_0 - 2i}{k_n(W/L)}}\right) - (2I_o - i)R + iR + 2\sqrt{\lambda}i_n R = 0 \tag{7}$$

And with regularization, following [8] (note noise term $[\sqrt{(2\lambda)}]R$, upon regularization, is multiplied by $\sqrt{\varepsilon}$, see eq. 4.12 of [8]):

$$\varepsilon \frac{di}{dt} = V - \left(\sqrt{\frac{2i}{k_n(W/L)}} - \sqrt{\frac{4I_0 - 2i}{k_n(W/L)}}\right) - (2I_o - i)R + \sqrt{\varepsilon}\left[2\sqrt{\lambda}i_n R\right] \tag{8}$$

Again, for simplification, for now we then select $I_0$, $R$, $(W/L)_{1-2}$ so that the coefficients have values of one. (5), (6) are then rewritten as:

$$\begin{aligned}\dot{x} &= f(x, y) \\ \varepsilon \dot{y} &= g(x, y) + \sqrt{\lambda \varepsilon}\eta\end{aligned} \tag{9}$$

---

[1] [7] uses eq. 18 of [7] (repeated here) to calculate jitter: jitter=(rms noise voltage in series with timing waveform)/(slope of waveform at triggering point). For the Schmitt trigger part it assumes the noise has no noise spike. With a fast pulse, the slope is large, so that the contribution of jitter is small. However simulations show similar results as Figure 8, where there are noise spikes (independent of how fast the pulse is), due to regenerative action. Thus even with large slope, the huge noise spikes, divided by this slope, still results in significant jitter.

$$\text{where } f(x,y)=y, \; g(x,y) = -x-y^3+y$$

(10)

Thermal noise is represented by white noise sources $\eta$, with intensity $\lambda$. The expression for noise term, given in (8), is also simplified, with the result expressed in $\lambda$.

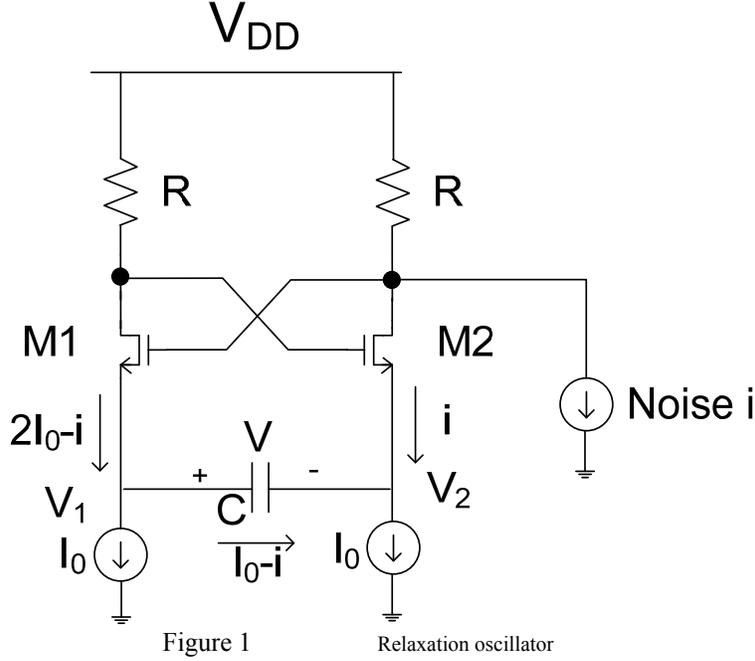

Figure 1  Relaxation oscillator

The noisy trajectories from $(x,y)$ describe trajectories in the x-y phase plane. An example simulation is shown in Figure 2a. It is known that when a relaxation oscillator changes state, it jumps and the trajectories spread out as they approach the jump point (bounded by the 2 solid curves), as predicted by model described in (4) (with noise added) [8]. Figure 2b, c show the output, noise vs $t$, where noise spikes are evident at jump points.

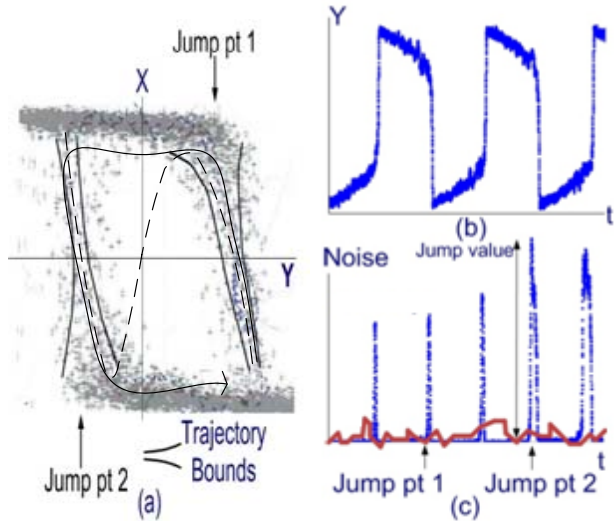

Figure 2  Matlab simulation, $\varepsilon=0.01$, $\lambda=0.05$ (a) phase plot : $x$ vs $y$, dotted line is $g(x,y)=0$ (cubic polynomial) in (6), Solid line with arrow is the deterministic trajectory (no noise). With noise added, the solid line got randomized. The resulting trajectories (plotted in discrete time) are represented as grey dots. Notice the bulk of the dots are within the two vertical tubes labeled trajectory bounds. Also notice the trajectory spread out towards the jump points, so that the bounds also spread out. (b) waveform : $y$ vs $t$ (c) noise in $y$ vs $t$. Simulations for grounded capacitor oscillator with Schmitt trigger (fig.16 of [7]) show similar noise spikes at jump points.

Thus instead of using rms noise, [8] uses maximum noise, typically occurring at jump point. The jitter, instead of being given as in eq. 18 of [7] (repeated here): rms noise voltage in series with timing waveform/slope of waveform at triggering point, is characterized by the ratio jump voltage/ slope of waveform at triggering point. To characterize this jump voltage, [8] defines trajectories $\Psi=(\Psi_x, \Psi_y)$, attributes a functional integral, $I_\varepsilon(\Psi)$, to it, and derives a probabilistic bound on the maximum noise:

$$I_\varepsilon(\psi) = \int_0^T \left( (\dot\psi_x - f(\psi_x,\psi_y))^2 + \left(\varepsilon(\dot\psi_y - \frac{1}{\varepsilon}g(\psi_x,\psi_y))\right)^2 \right) dt$$

$$\text{prob}\left\{ \max_{0 \leq t \leq T} (\psi_x - x) + (\psi_y - y) < \delta \right\} > \exp-\frac{I_\varepsilon(\psi) + h}{\lambda} \quad (11)$$

(12)

h is a constant larger than 0.
(12) is derived using large deviation theory where the exponential dependency arises from independence of white noise process, using the law of large number. However near the jump point, [8] admits that $i$ can have large deviation (after eq. 5.7 in [8]) and that (12) has severe drawback (remark 2 after eq. 5.5 in [8]). This is because the action $I_\varepsilon$ is an integral criterion, and the set that bound it is extremely badly behaved. As shown below we overcome this problem by mapping the problem into vdw gas. We then avoid having to go through the entire trajectory, evaluate the integral $I_\varepsilon$, in order to find fluctuation at the jump point. In this way we assume capacitor/resistor in oscillator/vdw gas is in thermal equilibirum. Thus we can employ the partition function concept. The fluctuation so calculated, can then be mapped back to relaxation oscillator.

## III. NOISE FOR OSCILLATOR (EQUATIONS SETUP AND QUALITATIVE DESCRIPTION)

We will set up the deterministic, then noise equations for oscillator and vdw gas. We will then start with noise equations and present the probabilistic behaviour of the resulting noisy trajectories.

### A. Deterministic:

1) *Setup equation of oscillator*

   i. *Introducing variational calculus/Lagrangian formulation*

In this section we follow [8] and review an alternative but equivalent way of solving the trajectory satisfying (5), (6), namely using variational calculus or Lagrangian (L) formulation [24] . This equivalence has been much in use in physics. The principle relies on solving the trajectory by considering all possible (and arbitrary) trajectories, evaluate the L associated with each of them, select the trajectory with the minimum L and concludes that to be the solution trajectory.

   ii. *A simple example: LC oscillator; consistent with direct solving*

As an example, a simpler version of the oscillator is presented. Here g(x, y) in (6) is simplified to be equal to –x and ε set to 1. Accordingly the circuit in Figure 1 is simplified to an LC oscillator. (5) becomes $\dot{x} = f(x,y) = y \quad \dot{y} = g(x,y) = -x$
Solving this directly gives the familiar solution: x=sin(t), y=cos(t)
Next we follow (11) and write out action for this oscillator, following [19], interpret the integrand as L, (Table 1, column 2, row 4):

$$L = (\dot{\psi}_x - f(\psi_x, \psi_y))^2 + (\dot{\psi}_y - g(\psi_x, \psi_y)) = (\dot{\psi}_x - \psi_y)^2 + (\dot{\psi}_y + \psi_x)^2$$

We apply the variational principle to L in the the Euler-Lagrangian equation: $\frac{\partial L}{\partial \psi_x} - \frac{d\left(\frac{\partial L}{\partial \dot{\psi}_x}\right)}{dt} = 0 \qquad \frac{\partial L}{\partial \psi_y} - \frac{d\left(\frac{\partial L}{\partial \dot{\psi}_y}\right)}{dt} = 0$

The $\Psi=(\Psi_x, \Psi_y)$ that satisfies the above equations are the deterministic x,y. Thus upon substituting L in these equations, and setting $\psi_x, \psi_y$ to x and y, respectively, we obtain the differential equation for the trajectory $\ddot{x} - 2\dot{y} - x = 0 \qquad \ddot{y} + 2\dot{x} - y = 0$
The solutions are x=sin(t), t*sin(t) and y=cos(t), t*cos(t)
As indicated in [8], trajectory $\Psi$ has to start at the same point as trajectories described by $\dot{x} = y$, $\dot{y} = -x$, at t=0. Hence the solutions t*sin(t), t*cos(t) are ruled out, and only sin(t), cos(t) qualify, which is consistent with solving directly.

   iii. *relaxation oscillator*

   a. *variational approach: consistent with direct solving*

We now apply variational calculus to oscillator. The detail is developed in Appendix I and summarized in Table 1. The trajectories are solutions to equations in (57), (58). As in the LC case, both (5), (6), and (57), (58), when solved, give the same trajectory, like that in Figure 2a. No closed form solution is available [56].

   b. *Direct solving : with coefficient evaluated*

Let us start with (5), (6) in section II and evaluate the coefficients in f and g explicity.
For (5), using (1) this becomes :

$$\frac{dV}{dt} = \frac{I_0 - i}{C}$$

(13)

For (6), using (2), after Taylor expanding the square root term, and keeping only up to 3rd order, this becomes:

$$-\varepsilon I_o dz/dt = V + \sqrt{8(I_o/g_m)}(43/64)(z) - (2I_o)Rz + \sqrt{8\, I_o/g_m}(1/64)(z)^3$$

(14)

Here normalized current

$$z = 1 - i/I_0$$

(15)

Dividing $I_o$ throughout, this becomes:

$$\varepsilon dz/dt = -V/I_o + \{-\sqrt{8(1/g_m)}(43/64)(z) + (2)Rz\} - \sqrt{8(1/g_m)}(1/64)(z)^3$$

(16)

and multiplying $g_m$ or

$$\varepsilon^* dz/dt = -Vg_m/I_o + \{-\sqrt{8}(43/64) + 2g_m R\}z - \sqrt{8}(1/64)(z)^3$$

(17)

where $\varepsilon^* = \varepsilon g_m$

If we assume parasitics only come from $C_{gs}$ in M2, then Appendix III shows that $\varepsilon = [(R/g_m)C_{gs}]$ and so

$$\varepsilon^* = RC_{gs}$$

(18)

2) *Setup equation of state of thermodynamic system exhibiting phase change; equation of state (with coefficient): vdw gas*

The variational principle above has been applied to physical system. The physical system we are presently interested in is the thermodynamics system, which as [9] has shown, can exhibit phase change, and which allows us to model jump behavior. The examples we are considering are the van der Waal (vdw) gas and Ising model for ferromagnet. Phase change in both cases are due to interaction between elements of system. In vdw gas, elements are the molecules and in Ising model these are the individual spins element on a lattice. The states of the molecules are characterized by their position and momentum, which are continuous variables while the states of the spin elements are characterized by the spin values, which are discrete. Because of this difference, different aspects of thermodynamic properties are more convenient to show using one model than the other and we will select the appropriate one as we go along.

To show the similarity of equation of state of the thermodynamic system to the corresponding equation of oscillator in (14), it is more convenient to use vdw gas, as the continuous nature of position/momentum variables are more compatible with the continuous nature of the current/voltage variables, used in the equation for oscillator. Thus following Appendix I.2), variables of trajectories of current/voltage in oscillator are mapped into pressure and volume of vdw gas. They are then minimized/maximized using variational calculus. Variable to minimize are Lagrangian in (54) (together with Rayleigh dissipation function $\Im$, obtained from dissipative term in (52)), and $L_{diss}$ in (59). Form of variational principle include Euler-Lagrangian (56), (57), (58) and Maxwell relations in footnote 11 in Appendix I. Equations of trajectories are described as solutions to (56), (57), (58). and equation of states as solutions to (60), (61), reproduced below:

$$dp/dt' = v$$

(19)

$$\varepsilon^{**} dv/dt' = g(p,v) = 4t' - p - 6t'v - (3/2)v^3$$

(20)

These have same forms as (13), (17). $\varepsilon^{**}$ is from 'fast' dynamics of the gas [8] and plays the role of $\varepsilon^*$ of relaxation oscillator in (18). Here 'reduced varaible', are used [34]. For example,

$$p = (P - P_c)/P_c, \quad v = (V - V_c)/V_c, \quad t' = (T - T_c)/T_c$$

(21)

These are just like $z = 1 - i/I_0$ in (15), where $V_c$ are like $I_0$, and are critical variables.

Special attention should be paid to t', the reduced variable for temperature T. The symbol t', is specifically selected for it, even though it is different from the time variable 't', because, according to Wick rotation [37], temperature variable T is the analog of time variable t. Thus we choose to use t' for the reduced temperature variable, to emphasize this relationship.[2]

---

[2] As in relaxation oscillator, where the differential equations can be obtained in parallel by writing KCL, KVL, here the differential equation can also be obtained in parallel from statistical mechanics method via the partition function[35].

Next, since we assume we operate near $T_C$, and since we are interested in fluctuation around phase change, t' does not change much. Therefore RHS of (20) is practically constant. Thus it has practically the same form as RHS of 2nd equation in (5), except now there is a vertical shift (due to 4t'), and the coefficient of the linear term is t', as opposed to 1, as it is normalized. The proper mapping will be revisited later, with normalization in (5), (6) lifted. Next, comparing (60), (61)to eq. 2.2, 2.5 of [9], we use t', the reduced temperature as independent variable, whereas[9] uses time t. Since f(t) in eq. 2.2 is rather arbitrary ([9] does not have specific constraint), we can take any f(t), as given in [9], and fits into our RHS of (60), (61), by specifying the necessary t'(reduced temperature) as a function t (time). Finally as shown in The detail is developed in Appendix I and summarized in Table 1. g(p,v)=0 (cubic polynomial) in (21) is plotted as dotted line in Figure 3. [3] This dotted line is similar to the dotted line for relaxation oscillator in Figure 2a. As expected the trajectory of p, v is expected to be like solid line in Figure 2a. where the trajectory p(t'), v(t') as t' progress, is similar to x(t), y(t), the trajectory of the relaxation oscillator, as time t progresses.

Finally the Ising model has similar behaviour (see for example, fig.12.14 [35], in the section on the Ising model in the zeroth approximation, for isotherms similar to Figure 3).

### B. Fluctuation (Noise)

The conclusion in deterministic section says that oscillator trajectory/electron trajectory(equation of motion)/vdw gas trajectory(equation of state) is derivable from the Hamiltonian. When we add the noise back, this trajectory has a random element at each point in the phase diagram, so that instead of one single trajectory, a set of (random) trajectories result. It turns out that the quantity that quantifies this randomness, depends on action/partition function, derivable from the Hamiltonian (part of path integral formulation).

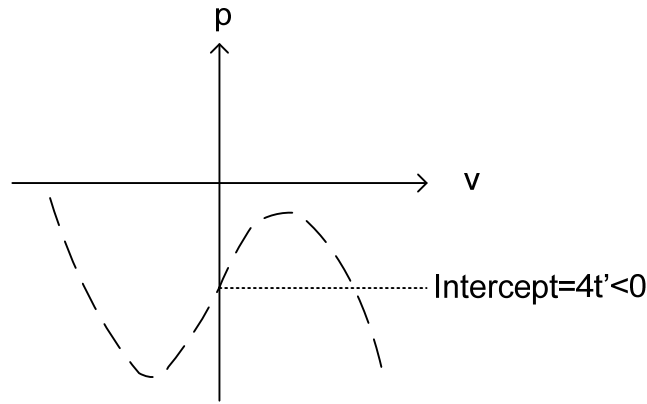

Figure 3    Dotted line is g(p,v)=0 (cubic polynomial) in (61) for vdw gas. This dotted line is similar to the dotted line for relaxation oscillator in Figure 2a. As expected the trajectory of p, v is expected to be like solid line in Figure 2a.

Before we pursue the variational viewpoint on noise further, let us introduce another concept useful in our noise investigation i.e. metastable state. We will reiterate qualitatively the jump/phase change mechanism and show the same mechanism happens in metastable state, which both the relaxation oscillator and the thermodynamic system exhibit. This, together with the variational viewpoint, will show the similar fluctuation mechanism between oscillator and thermodynamic system.

| | Table 1. Deterministic trajectory equation of motion/state | | |
|---|---|---|---|
| | Circuit (oscillator) | electron in magnetic/electric field | vdw gas/Ising model |
| Variable | current and voltage | position x and y | pressure and volume |
| parameter | time | time | temperature |
| Minimize/ Maximize | Lagrangian L | Lagrangian L | Entropy S |
| variational principle | Euler-Lagrangian | Euler-Lagrangian | Maxwell relations |

---

[8], [9], footnote 5, the trajectory p(t'), v(t') as t' progress, is similar to x(t), y(t), the solid curve in Figure 2a, in the present deterministic case i.e. no noise. When noise is added, there is difference in jump/phase change point. Nevertheless the fluctuation at jump/phase change point stays the same, as shown in section below.

[3] The Ising model has similar behaviour (see for example, fig.12.14 [35] in the section on the Ising model in the zeroth approximation, for isotherms similar to Figure 3).

1) *State switch/jump/phase change due to regeneration/symmetry breaking in a metastable state*

To link the noise in relaxation oscillator to vdw gas, let us start to make a few observations. First, from [36] an astable multivibrator, like a relaxation oscillator, can be constructed from a bistable multivibrator and an RC circuit. Again, from [36] a bistable multivibrator, consists of a positive feedback loop, and is capable of settling into two stable states. Such an astable multivibrator thus switches between these two states [4] periodically, just like the relaxation oscillator in Figure 1. In addition from [36], the bistable multivibrator can also exist in a metastable state. Fig. 17.18 of [36], redrawn in Figure 4a, likens this metastable state to be a form of unstable equilibrium, where the physical analog is a ball is the top of the hill (solid boundary), and any disturbance will cause the ball to one side or the other (dotted boundary) i.e. to a stable state at lower potential energy (oscillator analog will be voltage $V^+$ or $V^-$). Being in a positive feedback loop, noise in the circuit regenerates and got amplified. In the corresponding relaxation oscillator, during jumping (jump point 1, 2 in Figure 2c), regeneration also occurs, and the positive feedback loop also amplifies noise inside circuit ($i_n$ in Figure 1), creating noise spikes (noise spikes in Figure 2c).

Examples of metastability in circuit community can be found in [11], [12], [13], [14].
The mathematics of metastability is covered in [15], [18], [19],.
The physics of metastability is covered in [15], [16].

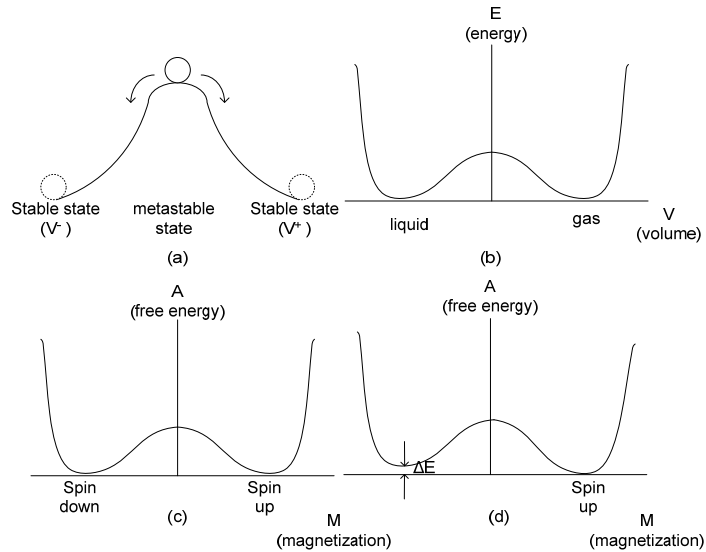

Figure 4     Phase change, metastable, broken symmetry

Second, as pointed out by [9], phase change in vdw gas is like jumping in relaxation oscillator. We would like to draw the analogy closer to the description in the above paragraph. In physics, there is the equivalent concept of metastability (pg. 339 of [15]) associated with phase change. Fig. 17.18 of [36], where with slight disturbance, the ball will roll to one side or the other (both at lower potential), manifests itself equivalently as in Figure 4b above [16]. Following [16], E is the internal energy of the vdw gas, and V the volume (like potential energy and position of ball, respectively, in Figure 4a). Again at the midpoint between gas and liquid, the vdw gas is metastable, and ready to do a phase change.

In this paper, for ease of illustration, we do one more mapping: vdw gas to Ising model (Ising magnets) [34]. Ising model is used to describe magnetization in a ferromagentic material. Magnetism comes from spin. Ising model describes the material under consideration as consisting of N spins arranged on a lattice. The energy of the system E, in a particular state v, with no external field, is [34]:

$$Ev = \sum(\text{energy due to interaction between spin})$$
$$= -J*\sum s_i s_j$$
(22)

Here spin $s_i = \pm 1$, J is a coupling constant, called exchange interaction (pg.394 of [35]). The magnetization M, is given as

$$M = \sum \mu s_i$$
(23)

μ=magnetic moment of the spin
With mean field approximation (22) becomes

$$Ev = -Jzm$$
(24)

---
[4] The relaxation oscillator, being an astable circuit, does not have two stable states, but rather two quasi-stable states [36], similar to the stable states in Figure 4a. It remains in each of these two states for a time interval determined by the time constant, as governed by (1), (2), and the threshold for regeneration.

Here z=number of nearest neighbour (to save notation we use z again, not to be confused with normalized current z in (15)), m=mean magnetization per spin. Jz is further denoted the exchange energy.

Figure 4b is redrawn in Figure 4c for the Ising model. Here A is the free energy of the Ising model, and M the magnetization, the volume (like potential energy and position of ball, respectively, in Figure 4a).

Now to go from metastable state to stable state $V^+$, in Figure 4a, Figure 4c is redrawn in Figure 4d, which shows the equivalent phase change to spin up. Notice the minimum on the left hand has its free energy raised up by $\Delta E$. Accordingly, the symmetry is broken (pg. 331 of [15] and fig.5.4 of [34]) and metastable state at the center changes into state on the right (spin up) state.

Moreover, just like relaxation oscillator, where noise regenerates and got amplified at jump point, similarly fluctuation in thermodynamic system also rises at phase change (pg. 129 [34]). Nevertheless, fluctuation in both cases are described by the same mathematics for fluctuation at the metastable state (see pg. 346[15], and thus is similar.

2) *Relation of fluctuation to variational approach.*

Let us now return to the variational approach and elaborates on using this framework to show how noise behavior is similar in both relaxation oscillator and thermodynamic system. We will then reiterates the previous discussion on metastable state to further connect the similarity.

The conclusion in deterministic section says that oscillator trajectory/electron trajectory(equation of motion)/vdw gas trajectory(equation of state) is derivable from the Hamiltonian. When we add the noise back, this trajectory has a random element at each point in the phase diagram, so that instead of one single trajectory, a set of (random) trajectories result. It turns out that the quantity that quantifies this randomness, depends on action/partition function, derivable from the Hamiltonian (part of path integral formulation).

With the variational principle this can be brought into more focus, since variational principle deals not only with deterministic trajectory (x y) in (5), but also noisy trajectory (x y) (9). Specifically [8] uses this approach on the system in Figure 1, whose Lagrangian is contained in the integrand of (11), which is further mapped into system describing an electron in a magnetic and electric field, described in (54) of Appendix I.1), to show the probability of a trajectory having a certain fluctuation, depends on action of the trajectory, given in (11), (12), and repeated in $2^{nd}$ column, $3^{rd}$ column of last 2 rows of Table 2. In the case of vdw gas, similar expression for system described in (59) of Appendix I.2) has also been derived, with action now determined by Lagrangian in (59). Again probability of a trajectory depends on action of the trajectory.

As shown in [21], [22], [23], the fluctuation mapping is first between thermal noise in relaxation oscillator to quantum fluctuation of electron in magnetic and electric field. Secondly, it is between quantum fluctuation in electron to thermal fluctuation in vdw gas. The conclusion is that, through the variational framework, fluctuation in relaxation oscillator is tied in to that in vdw gas, with probability of trajectory proportional to the exponent of negative of the trajectory action in each case, respectively.

However [8] shows, in the case of relaxation oscillator, to evaluate the fluctuation, this approach apparently runs into mathematical difficulty of badly behaved set(pg.662, remark 2 [8]), when applied to the relaxation oscillator. Essentially this is because of $\varepsilon$ going to 0. Similar situation arises in vdw gas, as $\varepsilon^{**}$ goes to 0.

Instead of looking at the entire trajectory, let us look at each point on the trajectory, which should remain mapped between relaxation oscillator and vdw gas. Now let us try to evaluate the fluctuation point by point. Let us try vdw gas. In doing evaluation point by point, let us further assume that when moving from one point to the other, the vdw gas is allowed to return to thermal equilibrium at each point. Under thermal equilibrium, the concept of ensemble is valid in describing the vdw gas [34]. Fluctuation in number of particles, $\delta N$, is given as:

$$\left\langle (\delta N)^2 \right\rangle = \sum_\nu N_\nu^2 P_\nu - \sum_\nu \sum_{\nu'} N_\nu N_{\nu'} P_\nu P_{\nu'}$$

(25)

Here N=number of gas particles, $\nu$ is the state, $N_\nu$ is the number of particles in state $\nu$. $P_\nu$ is the probability that system is in state $\nu$. For vdw gas it is described by grand canonical ensemble, with:

$$P_\nu = \exp(-\beta E_\nu + \mu N_\nu)$$

(26)

where $E_\nu$ is energy of the state and $\beta=1/kT$, $\mu$=chemical potential (to save notation we reuse $\mu$, used earlier in (23) as the magnetic moment). This is shown in $4^{th}$ column, last 2 rows of Table 2. Note that in thermal equilibrium, by neglecting dynamics [8], [9] $\varepsilon^{**}$ does not appear in the expression of $P_\nu$.

Now returning to the last section, from the similarity in metastable state viewpoint, the fluctuation, whether it be on relaxation oscillator, or vdw gas, at the jump point/phase change, remains governed by similar mechanism. We have also just shown that starting from variational framework, and specializing to looking at fluctuation at jump point/phase change, the mapping is valid. Moreover, we have a formula to evaluate such fluctuation in the vdw gas.

Our approach, therefore, is to solve for the fluctuation, for the vdw gas, using this "physics" approach, as opposed to the more "mathematical" approach in [8], [9]. The solution is then applied back to oscillator, by mapping the corresponding parameters. To map the parameters, we will set up the equations (deterministic) for relaxation oscillators and vdw gas. We will then compare and map the parameters. We will then solve for the fluctuation in vdw gas, map it back to relaxation oscillator. [5]

---

[5] Some comments on variational approach:

Comment 1: It is useful to note that the way we apply variational approach in deterministic case, one needs to know function f, g to describe the differential equation for the system. If one knows that, one could have solved for trajectory without going through variational approach. The reason for still doing this is because in the noisy

| | Circuit (oscillator) | electron in magnetic/electric field | vdw gas / Ising model |
|---|---|---|---|
| | | Table 2.   Noise properties of random trajectories | |
| Noise/fluctuation along trajectory | current | Position y | Volume/ susceptibility |
| Trajectory / Path/state characterized by: | Action I | Action I | Energy $E_v$ |
| Origin of randomness | Thermal noise | Quantum fluctuation | Thermal noise of gas molecule/spin |
| parameter for randomness | $\lambda=4kT/R+4kT(2/3)g_m$ | $\hbar$ =Planck constant | $1/\beta=kT$=thermal degree of freedom |
| Probability amplitude of each trajectory/path/state | $\exp(-I_\varepsilon/\lambda)$ | $\exp(-I/\hbar)$ | $\exp(-\beta E_v+\mu N_v)$ |
| Action $I_\varepsilon$ /Partition function $\Xi$ | $I_\varepsilon=\int \exp(-L_{diss}/\lambda)$ | $\int \exp(-I/\hbar)$ | $\Xi =\sum \exp(-\beta E_v+\mu N_v)$ |

3)  *Setup equation of oscillator (noisy)*

In the circuit picture, now let us add noise back, so that the circuit is described in Figure 1. Following (15), (16), and using noise term in (8), the circuit is described by the following:

$$\frac{dV}{dt}=\frac{I_0-i}{C}$$

(27)

$$\varepsilon dz/dt= -V/I_o+(1/I_o)\{-\sqrt{8}(I_o/g_m)(43/64)(z)+(2I_o)Rz\} -\sqrt{8}(1/g_m)(1/64)(z)^3 + \sqrt{\varepsilon}[2\sqrt{(\lambda)}i_n R]/I_0$$

(28)

Now at jump point, oscillator is at metastable state, and both M1, M2 are on. Hence the noise contribution comes from M1, M2 and the 2 resistors. We simplify by adding them together, so that power spectral density of 'noise i', denoted as psd of 'noise i', is given as ($\gamma$ is excess noise factor due to short channel effect, and using long channel approximation is 2/3)

$$\text{Psd of 'noise i'} =4kT/R+4kT/R+ 4kT\gamma g_{m1}+4kT\gamma g_{m2}$$

(29)

4)  *Setup of equation of state of vdw gas(noisy)*

(20) is the equation of average quantity only. Let us repeat and try to add fluctuation or "noise" to (20). Following [8], [9], this is written as (see eq. 2.7 of [9]):

$$dp/dt'=v$$

$$\varepsilon^{**}dv/dt=4t'-p-6t'v-(3/2) v^3+\sqrt{(\lambda \varepsilon^{**})}\eta$$

(30)

$\eta$ is noise source for volume fluctuation in vdw gas, $\lambda$ being the intensity of $\eta$ (to save notation we reuse $\eta$, $\lambda$ as in (9), but now refers to vdw gas rather than oscillator )and is [35]:

$$\lambda \text{ of vdw gas} =kT$$

(31)

---

case, the Lagrangian, action, also allows use to estimate fluctuation. In general in physics, when applying variational approach, the trajectory is not known a priori. The Lagrangian is obtained by L=T-U, where T and U are obtained from first principle. An example is the one in Appendix I.1)

Comment 2: [8], [9]  when using the varational approach, also show that   oscillator and vdw gas has different jump point, due to different order of taking ε→0 and λ→0. Nevertheless the fluctuation in both cases show the same exponential dependency on their respective action [21], [22]. Therefore the fluctuation we are going to derive for vdw gas, under thermal equilibrium, can be mapped to relaxation oscillator case. We should remember, as pointed out in footnote 12, that the mapping is only valid locally around the jump point, which is, after all, what we are interested here. We should also remember, in spite of the difference in jump point between oscillator and vdw gas, as discussed at the beginning of comment 2, nevertheless, as pointed out at III.B.1), the fluctuation is similar (and biggest) at the point of trajectory where the system is in metastable state  in both cases, which are the respective jump points.

In comparing (21) to (15), they are similar in that both are normalized, dimensionless variable. In the case of (15) it is $z=1-i/I_0$, and in (21) it is $v=V/V_c$. In thermodynamics v is the so called 'order' parameter, which characterizes fluctuation, and $V_c$, the critical variables. Likewise in circuit, i characterizes fluctuation.

5) *Fluctuation (Noise)*

In section III. B.1), it is explained that jump in relaxation oscillator corresponds to phase change, as the underlying cause in both case is due to breaking the symmetry in a metastable state. When comparing equations in oscillator to vdw gas, we now note that the change in slope of the cubic polynomial on RHS of (17) is responsible for jump in oscillator. Referring to Figure 2a, the cubic polynomial, $g(x,y)=0$, is the dotted line : there are 3 segments and 3 slopes; the center slope is positive and the two outer slope are negative. The center slope $dx/dy > 0$. This is condition of a limit cycle and jump (see pg. 260 [56], pg. 213 of [17]).

Similarly the change in slope of the cubic polynomial on RHS of (61) (i.e. $g(p,v)=0$) is reponsible for thermal instability. Specifically, from Figure 3 the center slope has $\left(\frac{\partial p}{\partial v}\right)_{t'} > 0$. According to eq. 9.2 of [16], this violates the stability criteria, and hence results in phase change in vdw gas (see pg. 44 of [34]).

In particular, the existence of a positive slope at the center of the i-v plot (Figure 2) for the oscillator, or equivalently at the center of the P-V plot for the vdw gas, is instrumental in the jump/phase change behavior. Adding noise, with the resulting equation in (21), does not change this conclusion.

Since slope is tied in with the coefficient, when we map between jump and phase change, we should compare these coefficients in the equations.

Thus the idea in this paper is to map the problem into vdw gas, and uses fluctuation obtained from there and map it back to relaxation oscillator by comparing coeffects.

To calculate fluctuation in vdw gas, one has to calculate its partition function.

Now from RHS of (61) one can calculate partition function, which depends on temperature T, or reduced temperature t'. This will be done in the following section.

## IV. FLUCTUATION FOR OSCILLATOR (QUANTITATIVE FORMULA)

We will start by calculating fluctuation formula in vdw gas. We will then map it back to relaxation oscillator.

### C. Vdw gas

1) *Before phase change.*

To investigate the fluctuation, we can start with (68). However a simpler way is to simply look at the mass transfer between the phases (liquid and gas) of a vdw gas. Before equilibrium we have [34]:

$$\Delta S = -(\mu_1/T - \mu_2/T) \Delta n \quad (32)$$

S is entropy. $\mu_1$ and $\mu_2$ are chemical potential of liquid and gas. $\Delta n$= net flow of particles between liquid and gas phase. $\Delta S>0$ and systems moves towards equilibrating.

At equilibrium $\Delta S=0$. Now since we are before phase change point, by definition $\mu_1 \neq \mu_2$. Substitute this in (97):

$$\Delta n = 0 \quad (33)$$

Thus mass transfer is practically zero. Thus one expects fluctuation to be small. A detail calculation using the exact formula from (25), (26), shows that fluctuation is indeed small [35] and is proportional to 1/V, i.e inversely proportional to volume, or size of the ensemble. This corresponds to the small noise in Figure 2c before the jump point. Specifically if for simplicity we consider the noise term is dominated by R and neglect the transistor noise (see footnote 6 below), then the noise on the capacitor/resistor system is inversely proportional to C, the equivalent of the size of the ensemble (it may be instructive to remember that in a simple RC circuit, noise is on the order of the familiar kT/C expression)

2) *At phase change.*

Unlike the case "before phase change", where $\Delta n=0$ in (33), here $\Delta n>0$.. To see why, again refer to (97).

Picking up from discussion after (97), which is for the case $\mu_1 \neq \mu_2$ (i.e. not phase change, no jump and $\Delta n=0$), now we are at phase jump, so $\mu_1=\mu_2$. Again equilibrating means $\Delta S=0$, but this time, with $\mu_1=\mu_2$, $\Delta n$ can be non zero and still satisfy (97).

This results in noise spikes. It corresponds to the noise spike in Figure 2c at the jump point.

How big is this jump? To answer this we revert back to the exact formula from (68), and applying (26) for grand canonical ensemble.

We follow and summarize Appendix II, where starting with (25) ((68) in Appendix II), we apply the concept of partition function $\Xi$ (3rd column, last row of Table 2) of the ensemble in thermal equilibrium. Using statistical mechanics principle we first write the mean $<N>$ in terms of $\Xi$: $<N>=\partial(\ln(\Xi))/\partial(\beta\mu)$ (this follows [34]). This is repeated to find the variance:

$$\left\langle (\delta N)^2 \right\rangle = [\partial^2 \ln \Xi / \partial(\beta\mu)^2]_{\beta,V}$$

(34)

(this follows (70), (71) of Appendix II, and repeated here)

Following (73), (76) of Appendix II, this is in turn expressed in terms of macroscopic quantity, $\beta$, P, $\rho$ as:

$$\left\langle (\delta N)^2 \right\rangle = [\partial \beta P / \partial(\rho)]^{-1}_{\beta,V}$$

(35)

Here $\rho$ is the density of vdw gas. Notice solving the right hand side involves expressing macroscopic quantity P in terms of $\rho$, i.e. V. This in turn involves the equation of state i.e. RHS of 2nd equation of (64) in Appendix II, without noise term, or equivalently RHS of (20), (61)

Following (69) this is given as $<\delta N^2>^{1/2}/\delta N$. Instead of $\delta N$, we choose $\delta\rho$ for the fluctuation, and so fluctuation is denoted as $<\delta\rho^2>^{1/2}/\delta\rho$. The final equation is (76), repeated here:

$$\frac{\left\langle (\delta\rho)^2 \right\rangle^{1/2}}{\rho} = \frac{1}{\sqrt{\rho V}} \frac{(1-b\rho)}{\sqrt{2\beta a\rho(1-b\rho)^2 - 1}}$$

(36)

Here "a" represents $\varepsilon_{interact}$, the scaling factor of interaction potential in (55). b is the size of molecule, and $\beta=1/kT$.
Near critical point, and for a specific case when

$\frac{\beta}{\beta_c} = \frac{27b}{8a}$  V=100b, $\rho=1/3b=\rho_c$. Then $\beta_c/\beta=kT/kT_c= T/T_c$, $\rho=\rho_c$. ($\beta_c$, $\rho_c$, $T_c$, $P_c$ are all critical quantities), we have, (36) of this paper becomes (this follows (78)):

$$\frac{\left\langle (\delta\rho)^2 \right\rangle^{1/2}}{\rho} = \frac{1}{5\sqrt{3(1/(\beta_c/\beta)-1)}}$$
$$= \frac{1}{5\sqrt{3(1/(T/T_c)-1)}}$$

(37)

For typical values of $T/T_c$, this fluctuation is quite big. It is inversely proportional to $\sqrt{(T_c/T-1)}$ and increases towards infinity as T approaches $T_c$.

If we define $w=T_c/T$, we can plot this in Figure 5

In general, when $\rho$ is not at critical value of $\rho_c$, we have similar form, except it does not peak up exactly at w=1. For example if $\rho=1/6b$ we have $\frac{\left\langle (\delta\rho)^2 \right\rangle^{1/2}}{\rho} = \frac{1}{5\sqrt{3((32/25)w-1)}}$. In general one can rewrite (36) as:

$$\frac{\left\langle (\delta\rho)^2 \right\rangle}{\rho^2} = \left(\frac{\left\langle (\delta\rho)^2 \right\rangle^{1/2}}{\rho}\right)^2 = \frac{1}{V}\left(\frac{1}{\sqrt{\rho}} \frac{(1-b\rho)}{\sqrt{2\beta a\rho(1-b\rho)^2 - 1}}\right)^2$$

(38)

However from [35] we also have:

$$\frac{\langle(\delta n)^2\rangle}{n^2} = \frac{1}{V}kT\chi_T$$

(39)

Where $\chi_T = -\frac{1}{V}\left(\frac{\partial V}{\partial P}\right)_{N,T}$ is the isothermal compressibility. Equating $\frac{\langle(\delta\rho)^2\rangle^2}{\rho^2}$ to $\frac{\langle(\delta n)^2\rangle}{n^2}$, we have

$$kT\chi_T = \left(\frac{1}{\sqrt{\rho}}\frac{(1-b\rho)}{\sqrt{2\beta a\rho(1-b\rho)^2 - 1}}\right)^2 \text{ or}$$

$$\chi_T = \frac{1}{kT}\left(\frac{1}{\sqrt{\rho}}\frac{(1-b\rho)}{\sqrt{2\beta a\rho(1-b\rho)^2 - 1}}\right)^2$$

(40)

Now at phase change (irrespective if one is near critical point) $\chi_T$ is quite big. This is because $\mu_1=\mu_2$, gas and liquid changes, but at practically a constant pressure, while volume changes (expands when going from liquid to gas, contracts vice versa). Thus $\left(\frac{\partial V}{\partial P}\right)_{N,T}$ goes to infinity. Meanwhile before phase change $\left(\frac{\partial V}{\partial P}\right)_{N,T}$ is small and so $\chi_T$ is small. Thus from (39) fluctuation is dominated by kT/V, or proportional to 1/V. This is consistent with section IV.C.1),i.

### D. Relaxation oscillator

We will now show a similar fluctuation formula to (37) can be developed for oscillator, by mapping coefficients between the equation of trajectory (in oscillator) and equation of state (in vdw gas). In particular we will concentrate on coefficients relating to interaction in both cases. We then try to come up with a "guess" of a reasonable fluctuation formula.

Next we will justify, physically, why these coefficients are related to interaction. In vdw gas, we have derived the fluctuation formula in last section and then from it identify that the underlying physical mechanism for fluctuation is interaction, which is also responsible for phase change. In oscillator we reverse the steps. We start with physically the reason for phase change, which is jump in oscillator, and identify the corresponding interaction mechanism.

#### 1) *Reason for having noise/fluctuation/regeneration*

Fluctuation is due to interaction. To identify interaction, we explore the phase change/jump process. We will now show that, like vdw gas, the reason of phase change/jump in relaxation oscillator is also due to "thermal energy" and "interaction potential" competes with one another. In the process of showing this competition, the *"thermal energy" or "noise mechanism"* and *"interaction potential" or "interaction mechanism"* in oscillator will become clear and identified with circuit parameter. Having identified that, we can use this identified circuit parameter in determining fluctuation in oscillator.

#### i. *Noise mechanism identification*

We look at (28), (30), and focus on their noise term. For (28), since in oscillator, we are interested in current fluctuation, we look at its psd of the 'noise i', which is given in (29). For design values in this paper, the noise is dominated by R, then psd of 'noise i'=8kT/R.

On the other hand from (30), since in vdw gas, we are interested in volume/density fluctuation, we look at the intensity of $\eta$, which is given in (31) and is $\lambda$ =kT.

Now psd formula in (29) is calculated for a 1-sided BW (positive side). For correspondence to intensity of vdw gas a 2 sided BW is used [28], and the density is divided by two (e.g. noise psd of a resistor, is now 2kT/R, instead of 4kT/R) and so psd for 'noise i'=4kT/R. Thus to map the psd of 'noise i' in oscillator to intensity of vdw gas, we have 4kT/R ↔ kT [6] i.e.

kT ↔ 4kT/R

(41)

---

[6] To be more complete, one can use (29). The conclusion of the paper stays the same.

Here ↔ is the symbol for "correspondence/mapping" between relaxation oscillator and vdw gas.

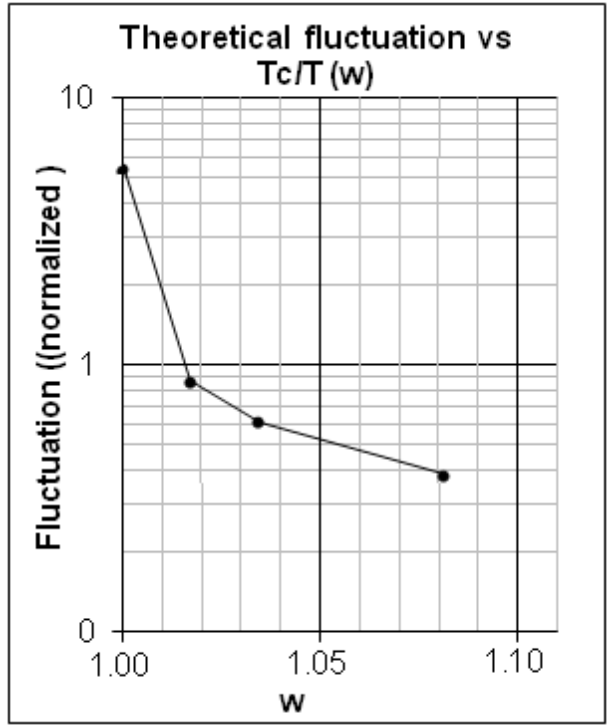

Figure 5        Theory: vdw gas. Fluctuation versus $w=T_c/T$.

*ii.   Interaction/deterministic mechanism identification*

*Mapping with normalization*

Since in section IIIB, we have established equation of oscillator (17),(28) are mapped onto equation of state of vdw gas (20), (30), we can compare the coefficients.

First let us go back to Table 1, where in row one (variable), we have P(pressure) ↔ V(voltage).

Next we can compare (28), (30), and concentrating on the coefficient of the linear term, for (28) we have:

$$\text{Coefficient of linear term} = \{-\sqrt{8(43/64)} + 2g_mR\})$$

(42)

This is roughly $2(g_mR-1)$.

Looking at (30) we have:
$$\text{Coefficient of linear term} = -6t = 6(1-T/T_c)$$

Thus $1/g_mR$ is compared to $T/T_c$ or $g_mR$ is compared to $T_c/T$

Looking again at the $T_c/T$ expression, which part is responsible for interaction/deterministic mechanism?
From noise mechanism identification we already note T as the factor responsible for noise mechanism in vdw gas (see, for example, left hand side (LHS) of (41)), thus the part responsible for interaction/deterministic is $T_c$..
Therefore in identifying the interaction/deterministic mechanism we should have $g_mR$ compared to $T_c$.
We now need to introduce the proper scaling factor. What are reasonable scaling factors?
Again let us go back to the relaxation oscillator equation that we used for comparison and look at the linear term i.e .linear term of (28). Instead of looking at the coefficient, we look directly at the variable i.e. z. From (15) z is dimensionless, being normalized by current $I_o$..

Now we go to (30) and again look at the variable of the linear term, which is v, or reduced volume. From (21), this is again dimensionless, being normalized by critical volume $V_c$, in a way similar to (15). From Table 1 row 1, volume corresponds to current. Hence we suggest a reasonable normalization constant is $I_o$, the current switched. With the corresponding voltage change as $\Delta V_{gs}$, the current switched can be expressed as $g_m \Delta V_{gs}$. A convenient form, is to follow [2], where the current switched, $I_o$, is shown to be proportional $g_m V_{peak}$. $V_{peak}$ is the voltage change (or swing) in V, the voltage across capacitor (shown in Figure 1) as current switches. Moreover interaction in vdw gas is on a per particle basis, we should also multiply by charge of the particle (electron or hole) in relexation oscillator, or q.

Since interaction in vdw gas is on a per particle basis, this should further be multiplied by charge of the particle in relexation oscillator, or q. Thus we have:

$$q g_m^2 R V_{peak} \leftrightarrow k T_c \tag{43}$$

### iii. Physical justification

How reasonable are (41), (43)?
(41) has been physically justified in a qualitative manner. (43), however, was arrived at mainly by comparing coefficients. In the following we will also has it physically justified in a qualitative manner.

### c. Justification by drawing analogy with fluctuation in circuit biased at unstable equilibrium/metastable circuit

First, to analyze fluctuation/noise of relaxation oscillator during jump, it is more convenient to look at the equivalent behaviour of a bistable circuit, as explained in section III.B.(1). We assume noise behaviour at jump and metastable point is similar, as regeneration [7] is responsible for the quick change of state in both cases (see [21] and Appendix IV).

Since we already know that both fluctuation and phase change of vdw gas arise from interaction, then we will show, in the next section i.e. section IV.D.(1).ii.d, that the equivalent state switching in astable circuit/relaxation oscillator, is physically from interaction parameter $q g_m^2 R V_{peak}$ i.e.

$$q g_m^2 R V_{peak} \leftrightarrow k T_c \text{ (see (47) below)}.$$

### d. Interaction parameter responsible for phase change/jump during state switching, in terms of circuit parameter

Following secIII.(1), fluctuation of relaxation oscillator when switching between quasi-stable states, is like fluctuation of the underlying bistable circuit when switching between stable states. In the case of bistable circuit, this switching of states happens when the circuit is at a metastable state.

The underlying bistable circuit operation is visualized in Figure 4a, where the two states states are $V^+$ and $V^-$, and the metastable state is 0V. This can be similar to a relaxation oscillator, with the mode of operation different. Referring to Figure 1, instead of having initially 0 volt at C, and M1 on, M2 one, we now have both M1, M2 on. Thus the circuit is fully symmetrical. V, the voltage across the capacitor, is therefore 0, just like the metastable state in Figure 4a. Again, it is in metastable state because even though it is stable as is, any tiny bit of noise is going to tip the symmetry to one side, with the regeneration (via positive feedback) kicking the circuit out of this metastable state, so that it settles into a stable state (either M1 on, M2 off, or vice versa).

### e. Circuit operation behind state switching in bistable circuit/relaxation oscillator and circuit parameter for interaction

Next, let us consider the moment of regeneration, when both M1, M2 are on. Initially V is 0, which then quickly settles to $V_{peak}$.
Let us first assume a charge q is deposited on the capacitor. The voltage change is q/C. The change in voltage is amplified by transistor M2 with load R, which has "gain" $g_m R$. Therefore its output (drain) voltage changes by $g_m R(q/C)$. This also changes the input (gate) voltage of M1 by $g_m R(q/C)$. Hence the current flowing through M1 changes by its transconductance multiplied by change in input voltage i.e. $g_m * g_m R(q/C)$. This change in current, is subtracted from current source $I_o$, which pulls the charge from bottom plate of C, also is responsible for pushing charge from M2 onto top capacitor. This should "flip/change" the next charge. Thus charge from M2 goes through regeneration and begets more charge.
Equivalently, as the change in voltage is amplified by transistor M2 with load R, which has "gain" $g_m R$. Therefore its output (drain) voltage changes by $g_m R(q/C)$. This also changes the input (gate) voltage of M1 by $g_m R(q/C)$. Again M1-R has "gain" $g_m R$. Therefore its output, or gate of M2 changes by a total of $g_m^2 R^2(q/C)$. This change in gate voltage pulls the source voltage of M2 by same value and flips the next $C*[g_m^2 R^2(q/C)]$ charge i.e. the next $g_m^2 R^2 q$ charge. Thus charge from M2 goes through regeneration and begets more charge. This is responsible for the interaction. To calculate the interaction energy, we note that such charges go across capacitor, with voltage taken to be $V_{peak}$ and so the interaction energy is $g_m^2 R^2 q V_{peak}$. Furthermore, since in section IV.D.(1).i, noise mechanism identification, there is a scaling factor of 1/R (see (41)) in the interaction mechanism identification, we apply the same scaling factor and so we have $g_m^2 R q^* V_{peak}$. The interaction mechanism for this state switching is equally valid in bistable and astable circuit (relaxation oscillator) and so we have:

$$q^* V_{peak} * g_m^2 R \leftrightarrow \text{interaction parameter}$$

(44)

*f. Comparison of interaction between bistable circuit/relaxation oscillator with thermodynamic system: Ising model*

Essentially we have two competing effects:
- The thermal "current energy/$\Delta f$" from $4kT/R$ randomize the polarity of the charge on the capacitor (can either flow in/out i.e. can be either positive or negative) and increases the entropy
- The regeneration circuit (M1, M2, R) supplies the interaction "current energy" to the charge and deposits charge of the same polarity. With same polarity, entropy decreases.

When comparing to thermodynamic system there are also two competing effects. Up to now, comparison of oscillator to thermodynamic system is performed using vdw gas as the example thermodynamic system. To show the similarity of interaction of the thermodynamic system to the corresponding interaction mechanism in oscillator, it is more convenient to use Ising model, as the discrete nature spin states (up/down) in demonstrating the positive feedback inherent in spin alignment/flipping and subsequent phase change, is more compatible with the oscillator, where the discrete nature of charge state (positive/negative) is used in demonstrating the positive feedback inherent in charge flipping and subsequent phase change.

Towards this end, we start by recounting that fluctuation depends on interaction. In general for thermodynamics system, $T_c$ depends on interaction. We will look at this relationship in Ising model.

From (22), J is the exchange interaction, which is related to $T_c$ as (pg.402 of [35]):

$$zJ = kT_c$$

(45)

z is the number of nearest neighbour to spin under consideration (to save notation, we reuse z, which is not to be confused with normalized current z as defined in (15)). Since $kT_c$'s unit is energy, zJ's unit is energy. As pointed out in (22), this represents the energy due to interaction between neighbouring spins, and [34] points out that such interaction, at phase change, propagates itself beyond nearest neighbour. The resulting long range correlation is what is responsible for phase change. It is also responsible for large susceptibility χ or fluctuation, at phase change. Thus interaction is the physics behind fluctuation.

Equivalently, for the relaxation oscillator, following [54], the charge on capacitor C itself can be modeled as an ensemble too. There the two conductors would have two different chemical potentials/Fermi potential. Refer to Figure 6a,b, for Ising model:
- Thermal energy kT randomize the spin
- The internal magnetic field Jzm supplies energy and acts to align the spin to have same sign. (zJ=exchange energy, m=mean magnetization per spin).

Just as zJ comes from internal magnetic field(molecular field) between neighboring spins, spatially, which is equivalently described by interaction potential between neighboring spins, $[\sqrt{2}-1)/2]*g_m^2 R(V_{gs}-V_t)|_{I=I_o}$ comes from regeneration circuit supplying energy to "neutralize then charge", individual particle, with charge q. Referring to Figure 6c,d, such charge, on a capacitor, with opposite polarity, can further be looked at like Ising magnets of opposite spin in Ising model. Now even though interaction is stronger among adjacent neighbors, however at regeneration/magnetization long range order exists and the underlying molecular field extends throughout i.e. it flips spins throughout. Similarly, the "analog" of the molecular field in oscillator, namely cross coupled transistors M1, M2, inject carriers, and changes polarity of charge throughout the top plate of the capacitor.

To elaborate on this resemblance in the "interaction potential" further, imagine the top plate voltage originally is positive at $V_{peak}$, as in Figure 6c (i.e. all put one charge is positive). This is equivalent to the case of Ising model when all but one spins are up Figure 6a. As negative charge is brought across capacitor, charge "2" flips from positive to zero, then to negative, with energy changing by one V (capacitor voltage) electron volt at every step (Figure 6d). This is equivalent to spin "2" is flipped from up to down, which means energy changes by one qJ. Next, with more negative charge brought across, then the next charge(spatially on the capacitor) is flipped to negative. Again in Ising model, the spin "3" flips from up to down. Hence it is seen that the spin flip plays the role of charge polarity changes, with the accompanying change in electron volt.

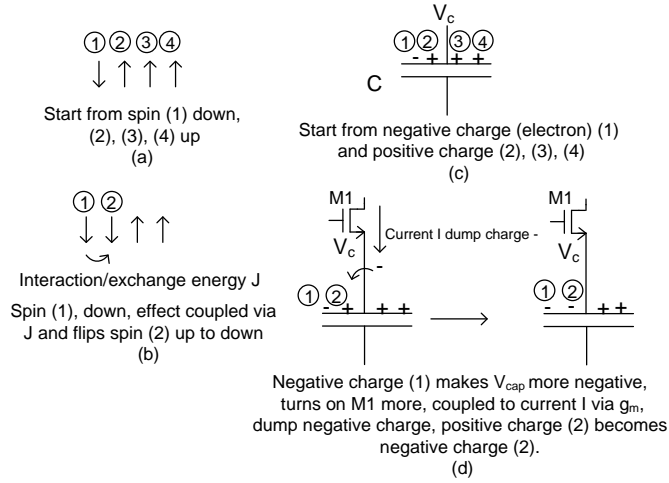

Figure 6  Comparison between spin flip (Ising model) and charge flip (capacitor in oscillator/bistable circuit)

In summary in oscillator, one can see charge is flipped(from positive/negative in time by transistor) on capacitor, making it resembling spin flipping.

***g.   Mapping***

We have identified the circuit parameter for interaction in relaxation oscillator. We have also justified, the similarity of the underlying physics, in interaction between oscillator and thermodynamic system, using Ising model as an example. The result, however, is general for thermodynamics system, be it Ising model or vdw gas. We can then map circuit parameter for interaction to physics parameter for interaction (derived in (45)) in vdw gas/Ising model. We summarize the steps as:

$$qg_m^2 RV_{peak} \leftrightarrow \text{interaction in oscillator}$$

(46)

Interaction energy in Ising model $\leftrightarrow zJ$

However from [35] $zJ = kT_c$

(47)

Thus $qg_m^2 RV_{peak} \leftrightarrow kT_c$

(48)

For a more quantitative physical explanation of (41), (43), (44) one can refer to Appendix IV, where an attempt is made to identify the partition function in the relaxation oscillator, and then justify (41), (43), (44).

In summary (48) is obtained, first through mapping coefficient of equation of state of thermodynamic system, using vdw gas as example. It is further justified, via mapping of interaction parameter of thermodynamic system, using Ising model as example.

2) *Formula*

To derive oscillator fluctuation formula, we note, following (36) for vdw gas, around critical condition, in general, for vdw gas [7]:

$$\frac{\langle (\delta\rho)^2 \rangle^{1/2}}{\rho} \propto \frac{1}{\sqrt{(1/(T/T_c)-1)}} = \frac{1}{\sqrt{(1/(kT/kT_c)-1)}} = \frac{1}{\sqrt{((kT_c/kT)-1)}}$$

(49)

---

[7] Comment 1: critical condition is elaborated in Appendix IV.
Commnet 2: a corresponding dependency on $T/T_c$, inside a square root function, is also true for Ising model. See pg. 431, eq. 29, in [35]

Now that we have identified the reason for peak up in vdw gas and the corresponding mechanism in the case of oscillator, as summarized in (41) and (48), we can use them to map (49).

Specifically we apply (41) and (48) to map $kT$, $kT_c$ to $4kT/R$ and $kT_c$ respectively, in (49). We also apply Table 1, which maps volume V to current i. This means density $\rho$ is also mapped to current i. With this mapping we have the following proportional relationship:

$$\frac{\langle (\delta i)^2 / \Delta f \rangle^{1/2}}{i} \propto \frac{1}{\sqrt{\left(\frac{qg_m^2 RV_{peak}}{4kT/R} - 1\right)}} = \frac{1}{\sqrt{\left((\frac{qV_{peak}}{4kT})(Rg_m)^2 - 1\right)}}$$

(50)

3) *Design Example*

An oscillator was designed with 0.13um CMOS. $V_{dd}$=1.5V. R=5k$\Omega$. W/L=0.5u/0.18u C=0.4pF. $I_0$=100uA. The simulated frequency is 55MHz. At room temperature, $qV_{peak}/4kT$ is about 1.
Then (50) becomes

$$\frac{\langle (\delta i)^2 / \Delta f \rangle^{1/2}}{i} \propto \frac{1}{\sqrt{\left((Rg_m)^2 - 1\right)}}$$

(51)

At $I_0$=95uA, $g_m R$=1.25 and at $I_0$=80uA $g_m R$=1. Now we define $\frac{qV_{peak}(g_m R)^2}{4kT}$ as the regeneration parameter. As we sweep $I_0$, $g_m$ and hence $g_m R$ sweeps, that is regeneration parameter sweeps. For a range of $g_m R$ from 1.0001 to 1.04, (corresponding regeneration parameter from 1.0002 to 1.09) the resulting noise at jump, i.e. $\frac{\langle (\delta i)^2 / \Delta f \rangle^{1/2}}{i}$, as given by (50), (51), is calculated. Essentially we are taking T to be fixed, and sweep $T_c$ (corresponding to oscillator with different design parameters such as $g_m$, R). The result is plotted in Figure 7. Here we plot noise vs regeneration parameter.

Note that like Figure 5, where fluctuation peaks up as $w=T/T_c$ approaches 1, here it peaks up as $(g_m R)^2$ approaches 1.

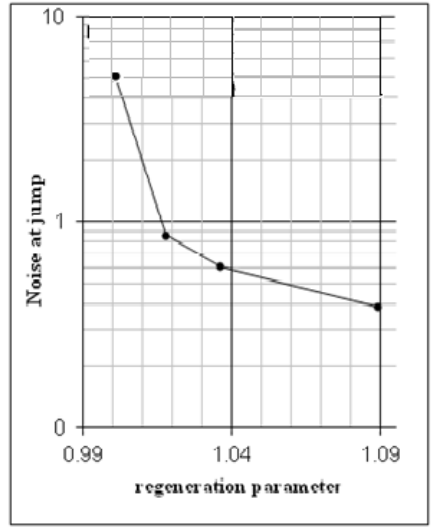

Figure 7    Theory: oscillator. Fluctuation versus regeneration parameter(=$T_c$/T)

V. SIMULATION RESULTS/EXPERIMENTAL RESULTS

We simulate a relaxation oscillator oscillator noise peaking as t approaches period (first done in Matlab simulation in Figure 2). This time we simulate a circuit designed in 0.18um CMOS. Shown in Figure 8 is Eldo simulation of the circuit in Figure 1 in 0.18µm CMOS, according to model described in (4) (with noise added) [8]. $f_0$=115MHz. Notice at jump point, there is noise spike.

Next simulation on thermodynamic system is performed. Because of the discrete states of spin variables in Ising model is compatible with the Metropolis method [34] (a Monte Carlo simulation), we choose to study the fluctuation of a thermodynamics system, using the Ising model as the example system.

The magnetization is calculated, as given in pg. 125 of [34] and from that the susceptibility χ evaluated as in pg.128 [34]. This is repeated by sweeping T, for a fixed $T_c$. If instead, we take T to be fixed, and sweep $T_c$, we would expect to get a similar plot, where, again, as $T_c$ (corresponding to different magnets with different $T_c$) reaches T, χ peaks up.

The resulting plot is shown below in Figure 9.

Compared to theory in both Figure 7, Figure 5, simulation in Figure 9, also shows fluctuation stays relatively small until the temperature/critical temperature ratio or regeneration parameter, $g_mR$ begins to approach 1 and there is a large increase in fluctuation or noise spikes [8].

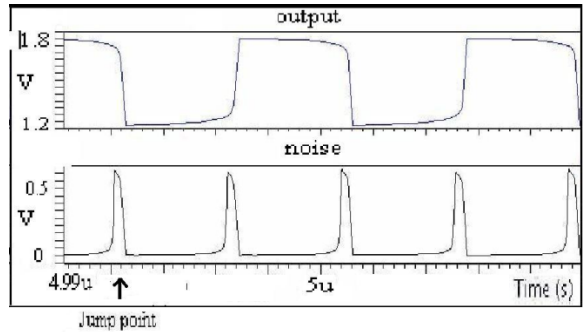

Figure 8    Circuit picture: Eldo simulation in 0.18um CMOS of Figure 1: $W/L$=1um/0.18um, $R$=3kΩ, $I_0$=100uA, $C$=3pF, $V_{dd}$=1.8V. $f_0$=115MHz, noise spikes at jump point.

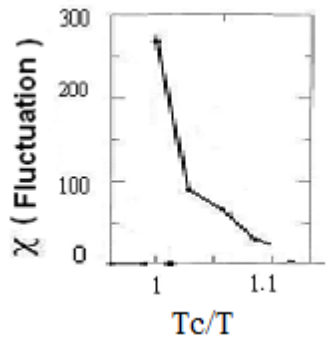

Figure 9  Simulation: Ising model. .Susceptibility χ (Fluctuation) versus $T_c$/T.

Next the design in section IV.C3) was fabricated and measured. The die photo is shown in Figure 10.
The measured output is shown in Figure 11, Figure 12, where frequency is 55MHz

---

[8] In simulation, fluctuation never rises up to infinity at regeneration point, due to numerical effect.
Figure 6 and Figure 5, 8 are prepared from a slightly different viewpoint. We can now reverse the viewpoint from discussion after (37), and state that for a given temperature T (say room temperature), if we select different gases (different $T_c$), fluctuation at phase change increases towards infinity if the selected gas $T_c$ is close to T, i.e. $kT_c$, interaction energy, is close to thermal energy kT.
This is perhaps more relevant in our case, where if we map thermodynamic system to oscillator, we typically operate at room temperature T. Then given this T, or thermal energy kT, if we show interaction energy $kT_c$, to depend on designs (W/L ratios, bias current $I_0$ etc.), then we can see how fluctuation varies, and increases towards infinity. That $kT_c$, or $T_c$, depends on design parameters (W/L ratios, bias current $I_0$ etc.), should not, perhaps, comes as a surprise. This is because, in vdw gas, as its $T_c$ depends on zJ, physical constants characterizing a particular ferromagnetic material, then in oscillator, its $T_c$ would be expected to depend on design parameters characterizing a particular oscillator.
Finally strictly speakin, for Ising model, plot should peak up at $T/T_c$=2.2 (i.e. shape remains the same but shifted to right so peak at 2.2); it is shown here to peak at 1 to avoid confusion with previous fluctuation plots, where serving to illustrate the general shape are same and that it peaks up at critical point.

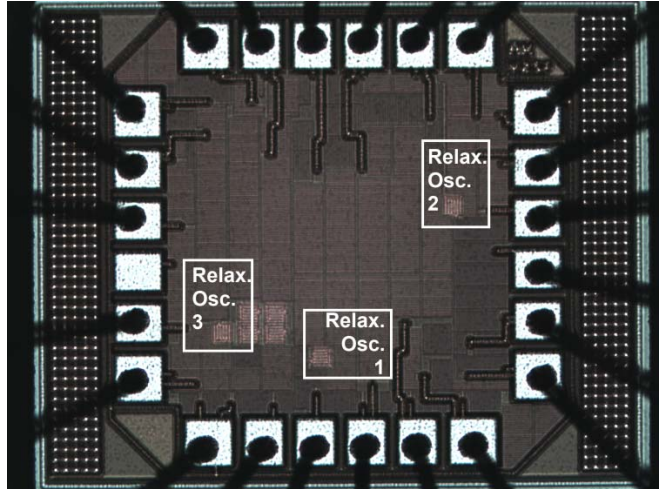

Figure 10                 Microphotograph of chip

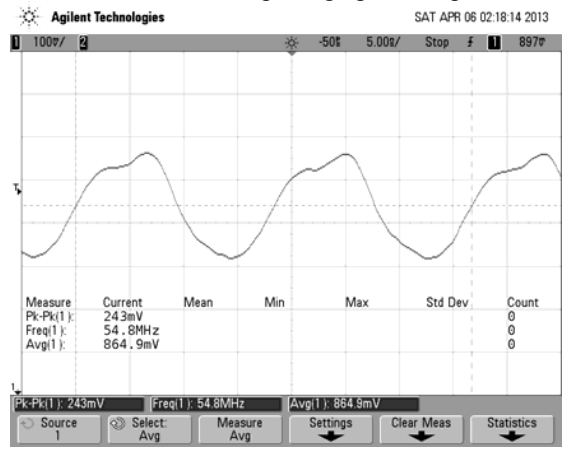

Figure 11                 Measured output, time domain

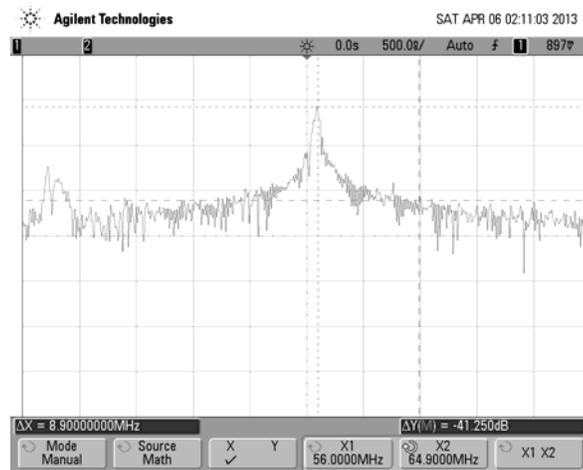

Figure 12                 Measured output, frequency domain, frequency 55MHz

Next to characterize the noise peak, instead of measuring the noise power directly, we measure the corresponding noise jitter, which is simpler to characterize [6]. Since noise spike happens at edge of transition, rms noise power directly translates into timing jitter [7]. To characterize the timing jitter, we use the phase noise module of the spectrum analyzer and measure the noise at 1MHz offset. Current $I_o$ is swept from 95uA down to 80uA, with corresponding change in $g_m$ and and $V_{gs}$-$V_t$, hence the regeneration parameter. Phase noise (PN) plots of two measurements at the two ends: 95A and 80uA, are shown in Figure 13.

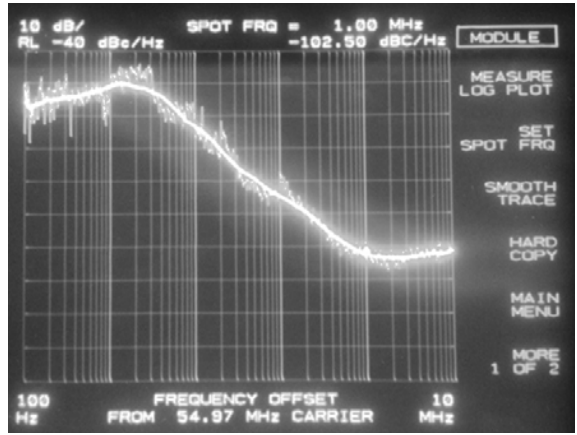

(a)

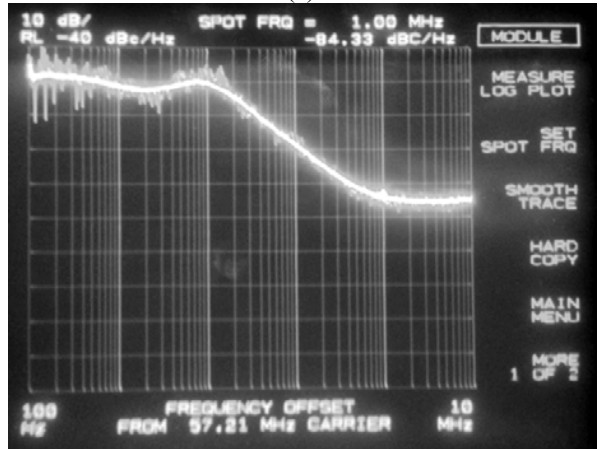

(b)

Figure 13 Measured phase noise plot (a) $I_o$=95uA. PN at 1MHz = -102dBc/Hz (b) at $I_o$=80uA. PN at 1MHz= -84dBc/Hz.

Note that as $I_o$ is lowered, $g_m$ and $V_{gs}$-$V_t$ are lowered, and the measured PN at 1MHz offset increases rather rapidly from -102dBc/Hz to -84dBc/Hz, as predicted by (51) and Figure 7.

Next phase noise at 1MHz offset from 4 values of $I_o$, between the two extremes, are taken. The phase noise are plotted against the corresponding regeneration parameter, in Figure 14. From these results we can show noise level increases as $g_m R$ decreases. The trend is like (51) and Figure 7. Note that as in Figure 7 and Figure 9, Figure 14 the noise rises sharply when the regeneration parameter, $g_m R$ (temperature/critical temperature ratio) approaches 1.

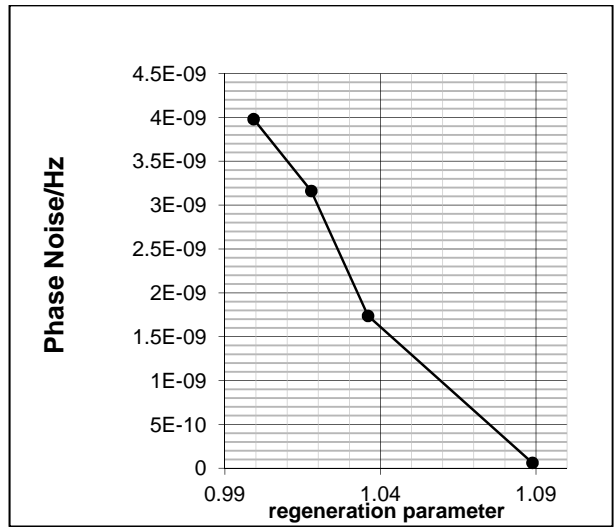

Figure 14    Plot for measured phase noise level vs regeneration parameter

## VI. CONCLUSIONS

For a relaxation oscillator, the regenerative nature of the circuit results in noise spikes. This paper attempts to explain this from a physics picture, using the partition function in thermodynamics. Using resemblance of noise spikes in regeneration in oscillator to fluctuation in phase change in thermodynamic system such as liquid/gas or magnetization, theory is developed. The mathematical tool used is the partition function in thermodynamics, and the results mapped between thermodynamic system and relaxation oscillator. Formulas were derived which show that the noise peaks up as the regeneration parameter, or loop gain, approaches 1. This is consistent with the expectation, as the loop gain of the positive feedback loop approaches one as $g_mR$ approaches one. Simulations on circuits (Eldo) using CMOS and Monte Carlo simulations (Metropolis) using Ising model, as well as measurement results on oscillator fabricated in 0.13um CMOS, show jump/spikes and the trend agree with theoretical prediction. Using the formula the designer can quantify the variation of noise spikes dependency on design parameters such as $g_m$ (device transconductance), R, $I_0$.

## ACKNOWLEDGEMENTS

The author would acknowledge help from D. Li, L. Bai, Stewart Robson, Forhad Hasnat, Prof. Afshordi, Nielsen, Heunis, Liu, Mann, McCourt, Roy, Wickham and Leggett, for pointing out [21] misses a dissipation term in its L.

## *Appendix I*

*Deterministic: equation of circuit oscillation/equation of motion of electron/equation of state of vdw gas*

1) *New Model in Electron in Electric and Magnetic Field :*

In this section, we will map the equation of oscillation of circuit to equation of motion of electron using variational calculus, thus justifying the Lagrangian description of the circuit. We start by looking at the case when oscillator is noiseless or deterministic. The trajectory corresponds to classical trajectory in the physics picture. The integrand of the action integral, for a conservative system, is interpreted as the Lagrangian. [24]. The present system has dissipation. The Lagrange's equations can be written in a form using Rayleigh function [24].

We start by expanding the integrand in (11) and identify the terms which are dissipative in nature (due to dissipation force), and setting $\Psi_y = \Psi_{y'}/\varepsilon$, we can write:

$$term\ dissipative\ in\ nature = \overbrace{+2\dot{\psi}_{y'}(\psi_{y'}/\varepsilon)^3 - 2\dot{\psi}_{y'}(\psi_{y'}/\varepsilon)}^{dissipative\ part}$$

(52)

Following [24] the dissipation force (e.g. friction) may be derived in terms of a function $\Im$, called the Rayleigh's dissipation function such that $\dfrac{d\ term\ dissipative\ in\ nature}{dt} = 2\Im$. Energy is dissipated, for example, in resistor R of Figure 1.

The rest of the term in the integrand is:

$$rest\ of\ term\ in\ \mathrm{int} egrand = \dot{\psi}_x^{\,2} + \dot{\psi}_{y'}^{\,2} + \left((\psi_{y'}/\varepsilon)^2\right) + \left(-\psi_x - (\psi_{y'}/\varepsilon)^3 + (\psi_{y'}/\varepsilon)\right)^2 - (2/\varepsilon)\psi_{y'}\dot{\psi}_x + 2\psi_x\dot{\psi}_{y'}$$

(53)

We now take these terms to give rise to the Lagrangian[9], in the sense that integrand of the action integral, for a conservative system, is interpreted as the Lagrangian:

$$\underbrace{\vec{L}}_{Lagrangian} = \overbrace{\dot{\psi}_x^{\,2} + \dot{\psi}_{y'}^{\,2}}^{T=KE} + \left\{ \overbrace{\left((\psi_{y'}/\varepsilon)^2\right) + \left(-\psi_x - (\psi_{y'}/\varepsilon)^3 + (\psi_{y'}/\varepsilon)\right)^2}^{pos\_dependent\ (electric)\ part\ of\ U} \right\} \overbrace{-(2/\varepsilon)\psi_{y'}\dot{\psi}_x + 2\psi_x\dot{\psi}_{y'}}^{vel\_dependent\ (magnetic)\ part\ of\ U}$$

(54)

With the term T, V, U identified, then L can be written as:

$$L = T - U$$

---

[9] An example is the LC oscillator described in IIIA1)ii. The corresponding physical example is again a charged particle in a uniform magnetic field in the z-direction, but the electric field expression is simpler. The solution of LC oscillator, sin(t), cos(t) is apparently consistent with the solution trajectory of the corresponding physical example of charged particle in uniform magnetic field.

$$\text{(55)}$$

In (54), (55) T=KE =kinetic energy, V=PE=position dependent potential, U=generalized potential, which includes V and the velocity dependent part due to magnetic force. Examples can include electromagnetic force, which is velocity dependent, but not dissipative.

Comparing to the *L* of a positive charged particle in a magnetic/electric field, with dissipation coming from collision from lattice potential, where we have: *L=T-U* , [27], with *T*=kinetic energy of the particle, *U*=potential energy due to electric field = -*q*φ, and magnetic field part = (*q/c*)*A.v*, *q*=electric charge, φ=electric potential due to electric field, *c*=speed of light, *A*=vector potential due to magnetic field in the z direction, *v*=velocity**,** dissipation part reflecting the nature of the lattice potential that introduces dissipation via scattering, Thus (54) can be interpreted as the L of such a physical system. Applying Euler-Lagrangian formulation (E-L):

$$\frac{d\left(\frac{\partial L}{\partial \dot{\psi}_x}\right)}{dt} - \frac{\partial L}{\partial \psi_x} = \frac{\partial \mathfrak{I}}{\partial \dot{\psi}_x}$$

$$\frac{d\left(\frac{\partial L}{\partial \dot{\psi}_{y'}}\right)}{dt} - \frac{\partial L}{\partial \psi_{y'}} = \frac{\partial \mathfrak{I}}{\partial \dot{\psi}_{y'}}$$

$$\text{(56)}$$

Substituting T, U, $\mathfrak{I}$ from (53), (54), (55) into (56), and setting $\psi_x$, $\psi_{y'}$ to x and y', we derive the classical equation of motion:

$$(\ddot{x}) - \left(1+\frac{1}{\varepsilon}\right)\dot{y}' - \left(-x - \left(\frac{y'}{\varepsilon}\right)^3 + \frac{y'}{\varepsilon}\right) = 0$$

$$\text{(57)}$$

$$(\ddot{y}') + \left(1+\frac{1}{\varepsilon}\right)\dot{x} - \frac{1}{\varepsilon}\left(\frac{y'}{\varepsilon}\right) - \frac{1}{\varepsilon}\left(-x - \left(\frac{y'}{\varepsilon}\right)^3 + \frac{y'}{\varepsilon}\right)\left(-3\left(\frac{y'}{\varepsilon}\right)^2 + 1\right) = \frac{1}{\varepsilon}\dot{y}'\left(3\left(\frac{y'}{\varepsilon}\right)^2 - 1\right)$$

$$\text{(58)}$$

(57), (58) are solved using the software Maple. x vs t, as plotted [10] in Figure 15, a periodic waveform is obtained, like the solution to (5), (6). **T**his shows the use of variational calculus/Lagrangian formulation in representing the circuit as a physical system is viable. Specifically the physical system is a single electron in a magnetic/electric field under lattice potential.

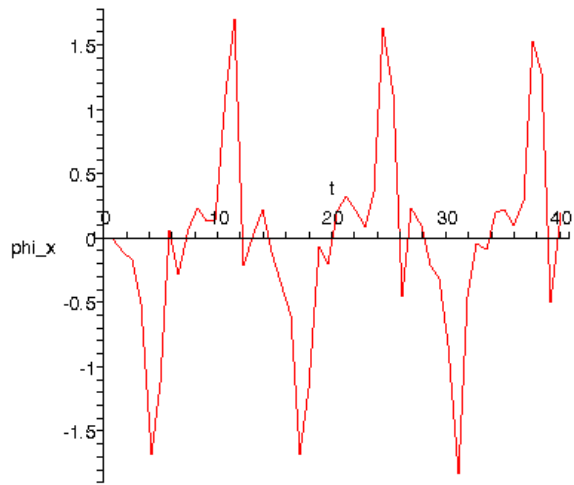

Figure 15          Physics picture: periodic solution to E-L equation, $\varepsilon=1$

---

[10] For illustration purpose, a simple value with ε=1 is used.

2) *New Model in Vdw Gas:*

$T_c$, $P_c$, $V_c$ are critical temperature, pressure and volume, respectively. $\dot{p} = \frac{dp}{dT}$  $\dot{v} = \frac{dv}{dT}$

We construct a similar dependency on p, v:

$$L_{diss} = (p-v)^2 + \left(\varepsilon^{**}(\dot{v} - \frac{1}{\varepsilon^{**}}\left(4t' - p - 6t'v - \frac{3}{2}v^3\right))\right)^2$$

(59)

With V'= $\varepsilon^{**}$V, (59) has similar form as (53)+(52) from the oscillator.

1) transform $L_{diss}(P, V') \to \hat{L}(\hat{P}, \hat{V}')$, which has no "dissipation term" via non-canonical transformation [31], [32]. Thus P, V $\to$ $\hat{P}$, $\hat{V}'$. Construct the corresponding $\hat{H}$

2) Since $\hat{H}$ obeys the canonical equation, and the entropy $\hat{S}$ also obeys the canonical equation, through Maxwell relation, then $\hat{S}$ forms the "analog" of the Hamiltonian $\hat{H}$ [11]. Thus trajectories $\hat{P}, \hat{V}'$ correspond [12] to trajectories of $\hat{\psi}_x, \hat{\psi}_{y'}$.

3) Upon transforming back, trajectories of P, V' correspond to trajectories of $\Psi_x, \Psi_{y'}$. Around critical point, we further have

dp/dt'=v

(60)

$\varepsilon^{**}$dv/dt'=g(p,v)=4t'-p-6t'v-(3/2) v$^3$

(61)

Here 'reduced varaible', t,v, p are used [34]. For example, t'=(T-$T_c$)/$T_c$ and is different from time t. $T_c$ is critical temperature. p=(P-$P_c$)/$P_c$, v=(V-$V_c$)/$V_c$.

$P_c$=a/27b$^2$   $V_c$=3b  $T_c$=8a/27bk

(62)

"a" is related to interaction potential, "b" is used to describe $r_o$ = radius of molecular volume or σ.

Similar situation applies to the ferromagnetic/paramagnetic example (when all the spins are lined up). Variational calculus can also be applied here, where again, the spin Hamiltonain, when subject to stationary phase, results in a solution for the equation of state [35]. This validates the use of variational calculus/Hamiltonian formulation in representing the circuit by a statistical mechanics system. [13]

---

[11] To recognize this analog, note that [24]: $\dot{x} = -\frac{\partial \hat{H}}{\partial \hat{p}}$  $\dot{p} = \frac{\partial \hat{H}}{\partial \hat{x}}$  ($\hat{P}$ = transformed momentum). The corresponding thermodynamics formulation is the

Maxwell relation: $-\left(\frac{\partial \hat{V}'}{\partial T}\right)_{P,n} = \left(\frac{\partial \hat{S}}{\partial \hat{P}}\right)_{T,n}$  $\left(\frac{\partial \hat{P}}{\partial T}\right)_{V,n} = \left(\frac{\partial \hat{S}}{\partial \hat{V}'}\right)_{T,n}$  ($\hat{P}$ = transformed pressure, $\hat{V}'$= transformed volume), together with the fact that $\hat{S}$ should be extremized (variation principle applied). Thus $\hat{S} \leftrightarrow \hat{H}$ and T$\leftrightarrow$ 1/k$\beta$ ($\beta$=1/kT) $\leftrightarrow$1/t in (52)-(54).

[12] To be exact, the analog is only true if we assume g function stays constant, as t changes i.e. as T changes in liquid/gas system. In reality, in the vdw gas, g, is also a function of T, and T is changing along the $\hat{P}$, $\hat{V}'$ trajectories. Here we are interested only in points on the trajectories around phase change/critical temperature point, $T_c$ i.e. not a period, but just a small local segment around jump point. Thus T is practically constant ($\cong$,$T_c$), and g is given by the reduced equation of state function (evaluated at the critical point $T_c$). In step 1, when we do the non-canonical transformation following [32], called rectification, we rely on assumption of local rectification i.e. T (and hence P, V') does not change much. As a side note, not only does [32] rely on P, V' not changing much, the equation there assume g is only a function of x, y, but not t. In the present case, the rectification method can be extended to varying t(or T) by using the non-autonomous extension[33].

[13] Note (61) is "deterministic" only in the sense that average value of P, V (in the trajectories) are taken. Obviously with thermal noise, T affects this trajectory(which shows up in coefficient of v). This is "like" biasing(dc) affects ac(noise) parameters.
We therefore are going to use the same formula as the deterministic part in the noise formula for vdw gas.

## Appendix II

*Fluctuation (Noise) Formula at Phase Change and Simulations in vdw gas*

The conclusion in deterministic section (Appendix I) says that oscillator trajectory/electron trajectory(equation of motion)/vdw gas trajectory(equation of state) is derivable from the Hamiltonian. When we add the noise back, these trajectories are randomized. It turns out that the quantity that quantifies this randomness, is derivable from action/partition function, which is intimately related to this Hamiltonian formulation.

First in the circuit picture, now let us add noise back, so that (16) becomes

$$\varepsilon dz/dt = -V/I_o + (1/I_o)\{-\sqrt{8(I_o/g_m)}(43/64)(z) + (2I_o)Rz\} - \sqrt{8(1/g_m)}(1/64)(z)^3 + i_n/I_0 \tag{63}$$

where $\lambda = 4kT/R + 4kT(2/3)g_m$

In [8] this is written as (9), (10).

1) *New Model in vdw gas:*

Now since P, V' in (60), (61) correspond to equation of state in a vdw gas, in any such trajectory, which we show in last section corresponds to equation of oscillation in relaxation oscillator, the fluctuation corresponds to the fluctuation in a relaxation oscillator.

Let us repeat and try to add "noise" to (61), just like adding noise to (16) to become (63).
Following [9], this is written as

$$\dot{p} = v + \sqrt{\mu}\xi$$
$$\varepsilon \dot{v} = 4t'\text{-}p - 6t'v - (3/2)v^3 + \sqrt{\lambda\varepsilon}\eta \tag{64}$$

Comparing (63), (64), the similarity between relaxation oscillator and vdw gas is observed. We are then led to conclude the noise behaviour around jump/phase change should be similar.

Next let us concentrate on the fact that we are specifically interested in fluctuation around jump. How do we find the corresponding fluctuation around phase change?

Specifically the phase change in P, V' diagram happens at the jump point. How do we calculate fluctuation at that point? We use partition function.

The partition function when the particle number remains fixed is canonical ensemble. This is given as $Q = \sum_\nu \exp(-\beta E_\nu)$. Since at the point of jump, there is phase change and the number of particles can vary, thus we generalize it to the grand canonical ensemble where the partition function $\Xi$, is given as:

$$\Xi = \sum_\nu \exp(-\beta E_\nu + \beta\mu N_\nu) \tag{65}$$

Here $\beta = 1/kT$, $\nu$, are the states, $E_\nu$, $N_\nu$ are the energy and number of particles in the state $\nu$. V is the volume, $\mu$ is the chemical potential. Its fluctuation reflects density $\rho$, and hence normalized current z, given in (21).

Furthermore, the probability of the state is proportional to the factor $\exp(-\beta(E_\nu + \mu N_\nu))$ ie.

$$P_\nu = \frac{\exp(-\beta E_\nu + \beta\mu N_\nu)}{\Xi} \tag{66}$$

Similar situation applies to the ferromagnetic/paramagnetic example (when all the spins are lined up).

In summary, remember we are calculating fluctuation, bounded by $\exp\text{-}I/\lambda$, at one point of oscillation. Due to difficulty of calculating I, we map (using the analog of entropy) and conclude fluctuation is same as in vdw gas. Then we further calculate fluctuation of vdw gas using partition function, which turns out to be path integral, but of $H_{micro}$(microscopic Hamiltonian of vdw gas). In the process of applying partition function to calculate fluctuation, we need to use the equation of state. Once we get the result we will map back.

2) *Fluctuation/Noise Formula: develop in vdw, map to relaxation oscillator*

Let us first define density $\rho$ as $\langle N \rangle/V$. Due to fluctuation, instantaneous value of the density differ from its average by $\delta\rho$. Hence fluctuation is $\langle \delta\rho^2 \rangle$ (i.e. the variance). What approach do we use to quantify this? From [34]:

$$\langle \delta\rho^2 \rangle^{1/2}/\delta\rho = \langle \delta N^2 \rangle^{1/2}/\delta N \tag{67}$$

Here $\delta N$ is the fluctuation in particle number. Now from

$$\left\langle (\delta N)^2 \right\rangle = \sum_\nu N_\nu^2 P_\nu - \sum_\nu \sum_{\nu'} N_\nu N_{\nu'} P_\nu P_{\nu'}$$

(68)

Using partition function[14], this can be expressed as:

$$\left\langle (\delta N)^2 \right\rangle = [\partial^2 \ln \Xi / \partial(\beta\mu)^2]_{\beta,V}$$

(69)

Here $\Xi$ is the partition function, $\beta=1/kT$, $\nu$, $\nu'$ are the states, $E_\nu$, $N_\nu$ are the energy and number of particles in the state $\nu$. V is the volume, $\mu$ is the chemical potential. From (65) $\Xi$ is given as:

$$\Xi = \sum_\nu \exp(-\beta E_\nu + \beta\mu N_\nu)$$

(70)

Since

$$<N> = [\partial \ln \Xi / \partial \beta\mu]_{\beta,V}$$

(71)

(68) can be written as

$$\left\langle (\delta N)^2 \right\rangle = [\partial <N> / \partial(\beta\mu)]_{\beta,V}$$

(72)

Using

$$\partial\rho / \partial(\beta\mu)]_{\beta,V} = \rho \partial\rho / \partial(\beta p)]_{\beta,V}$$

(73)

we have:

$$\left\langle (\delta N)^2 \right\rangle = \langle N \rangle [\partial \beta P / \partial(\rho)]^{-1}_{\beta,V}$$

(74)

What is $[\partial \beta P / \partial(\rho)]^{-1}_{\beta,V}$?

To answer this question, we need to use the equation of state. Using the partition function in (70), we can derive equation of state, for example by using Mayer cluster expansion and virial coefficient [35]. The final form, is

$$\beta P = \rho/(1-b\rho) - \beta a \rho^2$$

(75)

It is to be noted, near critical temperature, this becomes the familiar form given on the right hand side of (61). We can now differentiate this and get $[\partial \beta P / \partial(\rho)]^{-1}_{\beta,V}$ and substituting in (74) we have[15]:

$$\frac{\left\langle (\delta\rho)^2 \right\rangle^{1/2}}{\rho} = \frac{1}{\sqrt{\rho V}} \frac{(1-b\rho)}{\sqrt{2\beta a \rho (1-b\rho)^2 - 1}}$$

(76)

Here a, b are from vdw equation of state [9], with "a" represents $\varepsilon_{interact}$, the scaling factor of interaction potential. b is the size of molecule [35].

Near critical temperature this simplifies (where vdw equation of state (61) is indeed used). Using $\beta = \frac{1}{x}\frac{27b}{8a} = \frac{\beta_c}{x}$ V=100b, $\rho=1/3b=\rho_c$, we have:

$$\frac{\left\langle (\delta\rho)^2 \right\rangle^{1/2}}{\rho} = \frac{1}{5\sqrt{3(1-1/x)}} = \frac{1}{5\sqrt{3(1-T_c/T)}}$$

---

[14] Following [34], we make use of the relation $<N>=\partial(\ln(\Xi))/\partial(\beta\mu)$. The general form of this relation is $<X>=\sum P_\nu X_\nu =[\partial(\ln(\Xi))/\partial(-\xi)]_{\beta,Y}$, where X is any mechanical extensive variable, $P_\nu$ is probability of state $\nu$, $\xi$ is the conjugate set of a Legendre transform, $\beta=1/kT$, Y is all extensive variable not fluctuating.

[15] Actually we are using the full (not reduced) vdw equation of state

(77)

Here $x = \beta_c/\beta = kT/kT_c = T/T_c$ (note we are recycling variable x, which is different from x defined in (5)).
The above is for $T>T_c$, and for a fix $T_c$, with T changing. If instead, we take T to be fixed, and sweep $T_c$ (corresponding to different gases with different $T_c$), and starting from $T < T_c$, and changing $T_c$, we would expect to get a similar plot, with (77) becoming

$$\frac{\langle(\delta\rho)^2\rangle^{1/2}}{\rho} = \frac{1}{5\sqrt{3(1/(T/T_c)-1)}} = \frac{1}{5\sqrt{3(T_c/T)-1}}$$

(78)

Finally we can map the fluctuation formula for vdw gas in (78) to relaxation oscillation. This was done in the main text, in the development leading to (37), (42).

## *Appendix III*

*Example ε derivation for parasitics coming from $C_{gs}$*

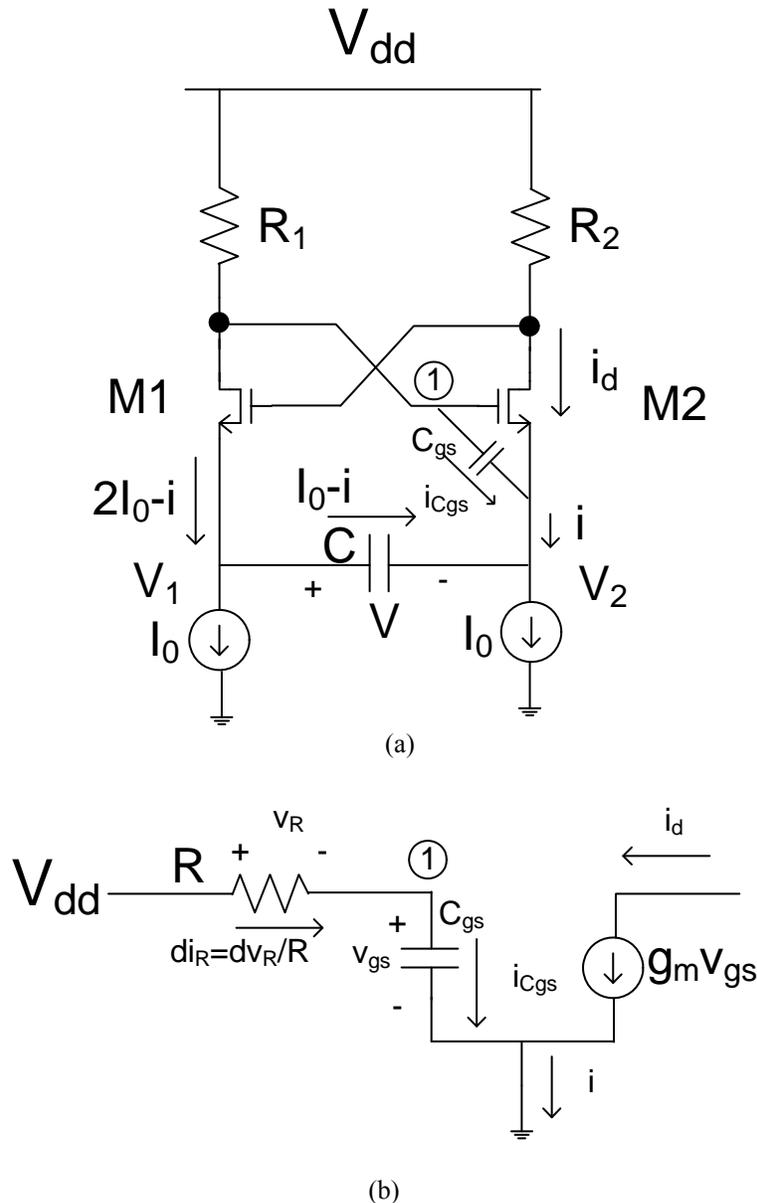

(b)

Figure 16    Circuit to calculate regularization parameter ε (a) parasitic $C_{gs}$ shown (b) small signal model of M2 including $C_{gs}$

The methodology is to observe the change in $V_{GS}$, $\delta(V_{GS})$, as a result of introducing parasitic capacitance $C_{gs}$. $\delta(V_{GS})$ is first related to ε via definition. The same change is independently evaluated via sensitivity analysis on the circuit, as a function of circuit parameters. The two results must be identical, and on comparing, ε is related to the circuit parameters.

Step 1: First, starting from eq. 4.1 of [8] (for BJT), we apply the MOS case, (4) above, and then we obtained δ(V$_{GS}$) as a function of ε, given below as:

$$\delta(V_{GS}) = -\varepsilon di/dt \tag{79}$$

Step 2: Second, we relate δ(V$_{GS}$) to circuit parameters via sensitivity analysis. We redraw Figure 1 as Figure 16a, with the addition of C$_{gs}$ of M2, assumed to be the only parasitics. M2, together with C$_{gs}$ is developed into its small signal mode, and shown in Figure 16b. The two resistors are labeled R$_1$, R$_2$ for explanation purposes but they have identical values R, as in Figure 1.

Referring to Figure 16a with C$_{gs}$ there is displacement current i$_{Cgs}$.

Referring to Figure 16b i$_{Cgs}$ means δi$_R$.

Meanwhile back to Figure 16a i$_{Cgs}$ means difference between i and drain current, i$_d$

Focusing on R$_1$ on left hand side, this also means additional current (=δi$_{R1}$), flowing through R$_1$.

This means additional voltage drop (=δv$_{R1}$), which is developed across R$_1$. δv$_{R1}$ is written as

$$\delta v_{R1} = R * \delta i_{R1} \tag{80}$$

Now applying KVL around the loop, starting from right hand side of C (at V$_2$): V + V$_{GS1}$ + V$_{R2}$ − V$_{R1}$ − V$_{GS2}$ = 0

If there is no C$_{gs}$, i.e. without additional voltage, repeating (2), we have:

$$V - \left(\sqrt{\frac{2i}{k_n(W/L)}} - \sqrt{\frac{4I_i - 2i}{k_n(W/L)}}\right) - (2I_o - 2i)R = 0 \tag{81}$$

With additional voltage drop across R (=δv$_{R1}$), this becomes:

$$V - \left(\sqrt{\frac{2i}{k_n(W/L)}} - \sqrt{\frac{4I_i - 2i}{k_n(W/L)}}\right) - (2I_o - 2i)R + \delta V_{R1} = 0 \tag{82}$$

However, from step 1, (79): the final expected form is δ(V$_{GS}$)= -εdi/dt, which, again by following [8], was given (4), repeated below:

$$\varepsilon \frac{di}{dt} = V - \left(\sqrt{\frac{2i}{k_n(W/L)}} - \sqrt{\frac{4I_i - 2i}{k_n(W/L)}}\right) - (2I_o - 2i)R \tag{83}$$

Comparing we have:

$$\delta(V_{R1}) = \varepsilon di/dt \tag{84}$$

Thus we want to relate δ(V$_{R1}$) to εdi/dt or δ(V$_{R1}$)/R= i$_{Cgs}$ to εdi/dt.

Referring to Figure 16a

$$2I_0 - i = i_{d1} \tag{85}$$

Differentiating (85)

$$-di/dt = di_{d1}/dt. \tag{86}$$

To find the relation between $\delta V_{R1}$ and $di/dt$ and then get $\varepsilon$, we need to find the relation between $\delta V_{R1}$ and $di_{d1}/dt$, , as in next paragraph.

Remembering, from (80),

$$\delta(V_{R1})/R = \delta i_{R1} = i_{Cgs} \tag{87}$$

We have from (87):

$$\delta(V_{DD}-V_{GS2})/R = \delta(V_{R1})/R = i_{Cgs} \tag{88}$$

Now to find $i_{Cgs}$ we note, from Figure 16b

$$i_{Cgs} = C_{gs} dv_{gs}/dt \tag{89}$$

However

$$C_{gs} dv_{gs}/dt = C_{gs} d(i_d/g_m)/dt \tag{90}$$

This is because, from Figure 16b

$$i_d = g_m v_{gs} \tag{91}$$

But we can further approximate this by noting:

$$i = i_{Cgs} + i_d \sim= i_d, \text{ as } i_d \gg i_{Cgs} \tag{92}$$

(which in turn is because $i_d = g_m v_{gs}$, while $i_{Cgs} = C_{gs} dv_{gs}/dt = \omega C_{gs} v_{gs}$, since $g_m \gg \omega C_{gs}$, for $\omega$ range of interest, which in turn because of parasitic is small i.e. $C_{gs} \to 0$ [16])

And so $i \sim= g_m v_{gs}$ or substituting this in (90) we have:

$$C_{gs} dv_{gs}/dt \sim= C_{gs} d(i/g_m)/dt \tag{93}$$

Substituting (89), (93) in (88)

$$\delta(V_{R1}) = \delta(V_{DD}-V_{GS2})/R = (C_{gs}/g_m) di/dt \text{ or}$$

$$\delta(V_{R1}) = \delta(V_{DD}-V_{GS2}) = [(R/g_m)C_{gs}] di/dt \tag{94}$$

Comparing (84), (94)

$$\varepsilon di/dt = [(R/g_m)C_{gs}] di/dt \tag{95}$$

so comparing coefficient

---

[16] Alternate (and perhaps simpler) reasoning: since no $\delta i_d$, $i_d$ is constant so $C_{gs} d(i_d/g_m)/dt$ (since $i_d = g_m v_{gs}$) ← $C_{gs} d(i/g_m)/dt$)

$$\varepsilon = [(R/g_m)C_{gs}]$$

(96)

## *Appendix IV*

*Partition function of relaxation oscillator*

To analyze this, let us reconfigure the relaxation oscillator so that while Figure 1 remains the same, for situation at Figure 6d , instead of having voltage equals maximum, now we have capacitor has zero charge, or capacitor voltage=0, while M1, M2 are both on i.e. on the verge of switching. This circuit we label as the metastable circuit in section IV.C.1) Of course, in this case the oscillator does not oscillate and M1, M2 just stay on indefinitely. In actual relaxation oscillator, the initial condition/voltage, V, is maximum (capacitor is charged) at time when M1, M2 are both on to maintain oscillation. We will assume noise behaviour in this metastable circuit carries over to the actual oscillator.

Now let us look at the metastable circuit noise behaviour at T below $T_c$. T below $T_c$ (i.e. interaction energy due to regeneration is big enough to overcome thermal noise so that it is possible to break symmetry and switch state)

1) we run with zero external field i.e. with no initial (initial means when M1, M2 both on) voltage/field across the capacitor, and with a small external initial voltage/field (much smaller than maximum voltage). In the zero external field, (of course, as stated above in this situation the metastable circuit does not oscillate; the capacitor voltage is identically zero, M1, M2 stay on and have equal strength, system is at an unstable equilibrium and can stay on indefinitely; due to randomness eventually it can be pushed to either state since interaction energy/feedback strength due to regeneration is big enough to overcome thermal noise), there are large noise fluctuation, due to correlation in fluctuation coming from interaction energy of the positive feedback loop (both M1, M2 on). [17]

2) With a small external field, symmetry is broken, state is switched and there is little fluctuation (little interaction as either M1, or M2 is off and feedback loop broken). [18]

As stated above, in the actual operation of relaxation oscillation, V is set initially to maximum so it oscillates. Furthermore, the interaction energy/regeneration is set strong enough i.e. we have enough loop gain, so that case (2) above is avoided. Under this situation the oscillator starts in one state and charges (like point (1)(b) above), switches states (during which it is like case (1)(a) above), discharges (like case (1)(b) above), switches states again (to opposite state, during which it is again like case (1)(a) above). Thus during state switching, like case (1)(a) above, it has large noise fluctuation (of course, when oscillator has settled in one state(charging/discharging) like case (1)(b) above, there is little fluctuation).

At $T<T_c$ i.e. the case when the metastable circuit switches state. This is equivalent to Ising and/or vdw gas below $T_c$, where there are distinct phases, due to broken symmetry of the bistable potential (see fig.5.4 pg. 126 [34]). Equivalently in metastable circuit, there is broken symmetry at case below $T_c$ i.e. when temperature is low enough that the corresponding thermal energy, kT, can be overcome by the interaction energy due to loop gain of positive feedback, and a state can be established. During switching there can be large noise fluctuation.

What is the equivalent situation in Ising model/vdw gas? Again it is easier to compare with the metastable circuit, and then map it to actual relaxation oscillator.

In Ising model and/or vdw gas, with broken symmetry (see fig.5.4 pg. 126 [34]), the surface energy does not go to zero, or equivalently the resulting surface tension keeps a continuous surface(see fig.5.7a-c pg. 132 [34]), with 2 distinct phases, or a single liquid droplet.

Let us now map back to actual relaxation oscillation. With $T < T_c$, here, the oscillator starts in one state and charges (like case (2) above), switches states (during which it is like (1) above), discharges (like case (2) above), switches states again (to opposite state, during which it is again like case (1) above). Thus during state switching, like case (1) above, it has large noise fluctuation (of course, when oscillator is in one state(charging/discharging) like case (2) above, there is little fluctuation). As stated above, this noise fluctuation is also due to long range correlation (regeneration/positive feedback), as in $T=T_c$ case.

To summarize the actual relaxation oscillator, the metastable circuit and Ising model/vdw gas all have similar noise behaviour. They all arise from long range correlation/regeneration/positive feedback. Next we quantify this noise behaviour by finding the partition function of the reconfigured relaxation oscillator. With partition function, we can apply that to find noise behaviour, just like in Appendix II, we apply the partition function of vdw gas to find its noise behaviour.

To find partition function one could have started by identifying the ensemble in relaxation oscillator. Let us take the species of ensemble as packet of net charge on capacitor C in Figure 1.

The capacitor is assumed to be metal plate capacitor. The ensemble consists of net charges.

---

[17] In physics, this corresponds to [34], pg. 129, where due to randomness the Ising magnet can be in either state, with χ changing by $2Nm_0^2\mu^2$, which blowup as N increases. This big fluctuation or χ comes from long range correlation.

[18] In physics, this corresponds to [34], pg. 131, fig.5.7c, where fluctuation is quenched when an external gravitational field is applied, and there is no longer long range correlation.

Initially the capacitor has zero net charge. These net charges in the top plate (RHS of C) are in thermal equilibrium. The chemical potential are $\mu_1$ and $\mu_2$, or Fermi potential $E_{ft}$ are $E_{fb}$ [54]. Initially with zero net charge, the voltage V is zero, and they are equal i.e. $\mu_1 = \mu_2 = \mu$, $E_{ft} = E_{fb} = E_f$. The other parameter is current i associated with C.

The capacitor is going to be charged. Let us assume the charge are positive charges. With net charge, there is voltage V on the plate. Eventually, C is charged so that M2 is in off state, V becomes $V_{peak}$.

This state is also not stable, the "inertia" is going to make the oscillator reverses and M1 starts discharging C, until M1 is in the off state.

The instant when M2 just hits off state, ( or $V = V_{peak}$) defines the beginning of jump point. During the jump, C discharges, and half way, we have again no charge on C.

At the beginning of jumping instant (M2 just off), we note that the top plate, is approaching when we have equal number of positive charges/negative charges. Because of regeneration, current i dumps the charge so that voltage drops approaches zero, in much the same way that phase change in Ising model is happening due to regeneration. At the jump/phase change, net charge is zero, and capacitor's voltage change sign, i.e. like vdw changes phase. As in vdw gas, approaching this phase change, each phase's particle number is not constant (e.g. gas transferring to liquid), where in present case, number of positive charges can change. Just to emphasize, in Ising/vdw and metastable circuit, this regeneration occurs as in case (1)(a) in pg. 23 ($T<T_c$), where state is changed due to small random external field, This small random external field is enough to break the symmetry/helps system to rolloff the unstable equilibrium point and switch the system. Large fluctuation is due to regeneration/correlation. In the actual relaxation oscillator, as stated above, because of initial voltage, the inertia will push the system to rolloff the unstable point. Nevertheless, similar regeneration/correlation happens, even though the rolloff is from inertia rather than some randomness and fluctuation again is big.

Now consider this duration of charging until M2 is off and denote it $T_{dura}$.

Consider in $T_{dura}$, the packet/domain of positive charge is Q. This Q is exactly positive charge on top plate at the instant when M2 is in the off state. Thus $Q=CV_{peak}$.

Now let us consider that during $T_{dura}$, this Q is in thermal equilibrium among itself. As the jump happens from M2 off to M1 off, there is positive, then positive, then negative charge on top plate.

Refer to Figure 6:
1) Assume the charges (positive charges/negative charges), are induced by stimulus (M1, M2 R) on top(right) place of C, form the system.
2) Initially Figure 1, M1 off, M2 on, $V_{peak}$ positive.
   i-$I_0$ is charging C, so that drain of M2 becomes more negative.
   Thus there is positive charge, i.e. and system not in equilibrium
3) As it approaches jump/phase change, which we interpret at beginning of this appendix, negative charges are brought in (from M2) so that net charge=0. We assume the charges go into a phase change.

   To elaborate on this, let us return to our model of the charge on a metal capacitor as an ensemble. Again we have charge on the top and bottom plate, with corresponding Fermi potential $E_{ft}$, $E_{fb}$, respectively. With C alone (i.e. no external circuit such as M1, M2, R), when there is zero charge, resulting in $E_{ft}=E_{fb}$, we would not have phase change, since as pointed out in [54], the two plates share energy, but not particles. However, in the present case, the two plates are connected, separately, via the cross coupled pair circuit, which act to transfer particles, in a "virtual" fashion. Following the discussion in section Appendix IV.C1) on *"Justification by drawing analogy with fluctuation in metastable"*, this virtual circuit, at moment of regeneration (or phase change), when V=0, i.e. $E_{ft}=E_{fb}$, a tiny bit of noise is going to kick this into regeneration mode. As each charge leaves R and enters transistor M2, the transistor gives energy to this charge and deposits the charge on the capacitor. The charge, by changing the voltage of V, sets up a "chain" reaction. The change in V is amplified by M2, then changes the gate of M1, then changes the current of M1. This change, when subtracted from constant current source $I_o$, pulls the charge from bottom plate of C. Therefore there is transfer of charge ("virtual") from top plate to bottom plate via this cross coupled pair. Note again the similarity to Ising model, where spin does not move/transfer, but it affects/flips the neighbouring spin, as if the spin has "transferred". One can see this happening in Figure 6a.

   Looking at this in another way, note even though looking at just the capacitor itself, charge transfer is due to charging(electrostatic), but looking at capacitor together with M1, M2, R, then there is not just charging, as there also is regenerative/feedback, and so the charge transfer is due to phase change mechanism, where charge transfer is like mass transfer in vdw gas phase change (remember in phase change, huge mass transfer at phase change point, from gas → liquid); in other words, not accountable by simply electrostatic charging.

   To investigate this mass transfer further, let us start before phase change. Now we have
   $$\Delta S = -(\mu_1/T - \mu_2/T) \Delta n \tag{97}$$

   $\Delta n$= net charge. $\Delta S>0$ and systems moves towards equilibrating. Then $\mu_1 > \mu_2$, $\Delta S>0$ (from 2$^{nd}$ law, as system equilibrates) means $\Delta n<0$. This flow of positive charges to negative charge is going to change $\mu_1$, $\mu_2$ until they are same, at which point, we have equilibrated, so $\Delta S=0$ i.e. this implies $\mu_1 = \mu_2$.

   Referring again to (97), right at the beginning, $E_{ft}>E_{fb}$ (in (97) is $\mu_1 > \mu_2$), and so $\Delta S>0$ towards equilibrating. Then $E_{ft}>E_{fb}$ (in (97) this is $\mu_1 > \mu_2$) which means $\Delta n<0$. Thus there is flow (virtual) of negative charge from top plate (leaving positive charge) to bottom plate. This charge buildup changes the electrostatic potential, and hence Fermi potential $E_{ft}$, $E_{fb}$ (in (97) is $\mu_1$, $\mu_2$) until they are the same. At this point we have equilibrated, so $\Delta S=0$ i.e. this implies $E_{ft}=E_{fb}$ (in (97) is $\mu_1 = \mu_2$) and we are on the verge of phase change (in vdw, this means we are on the verge of changing gas to liquid or vice versa; in relaxation oscillator circuit that means we are on the verge of changing from state M1 on/M2 off to M1 off/M2 on). To repeat, then,

4) phase change implies $E_{ft}=E_{fb}$ and $\Delta n>0$ (i.e. can fluctuate, or charge on top/bot plate can fluctuate) . Once regeneration ends (M1 off, M2 on), there is no more transfer of charge and $\Delta n=0$. Of course $E_{ft}<E_{fb}$(in (97) is $\mu_1<\mu_2$). Again $\Delta S=0$, which means we have again equilibrated. Of course in this case, the equilibrium is the state for (M1 off, M2 on).
5) With this regeneration, M1, M2 change state. Then the charges (positive charges/negative charges), which are induced by stimulus (M1, M2 R), form the system. Note R plays the role reservoir, which takes heat away from net charge. Thus referring to Figure 1, M1 on, M2 off, V negative. i-$I_0$ is discharging C, so that drain of M2 becomes more negative. Thus there is negative charge, i.e. and system not in equilibrium.

With the above description we can see the ensemble is characterized by a grand canonical partition function, because the number of positive charges can change, in much the same way in a vdw gas, the number of gas particle can change (it can condense into liquid).

Furthermore we can talk about the partition of the positive charge "phase", in much the same way as the partition function of the gas phase in vdw.

So far we have been considering partition function $\Xi$ of ensemble (of positive charge Q in oscillator) , which has same form as $\Xi$ **of** gas in vdw, as in (65), i.e. $\Xi = \sum_v \exp(-\beta E_v + \beta \mu N_v)$, where v is the state and sum is over all state.

The form is $\Xi$ ( $\mu$, V, T). (pg. 102, eq. 15 [35] )

However remember Q, which is exactly ensemble of charge on top plate when M2 just hits off, is also "packet of charge charged in duration $T_{dura}$". Therefore we can talk about "partition function" $\Xi$ of Q, as viewed as being charged during duration $T_{dura}$, as well. Our next goal is to establish the form of $\Xi$ for charge Q charged in $T_{dura}$.

$\mu$ =?

For ensemble of positive charges=Ensemble of charges on top plate when M2 just hits off we have $\mu$:

$$E_{ft} = E_{fb}, \text{ which}= \mu \tag{98}$$

V=Volume=? Or equivalently $\rho$ =density=?

Start with Figure 1 and focus on C. Let us remove the rest of the circuitry. With the standalone C, charge q on the plates are not interacting and C just behave our a container of charge, and it obeys the equation Q=C*Voltage. An analogy can be found in a container of ideal gas (again no interaction between gas particles), where it obeys equation P*Volume=NRT, (N=number of moles) or since density=$\rho$=N/volume, we have $\rho \propto$ P. Comparing with Q $\propto$ voltage, we identify $\rho \leftrightarrow$ Q [19]. Now back in Figure 1 and with the rest of the circuitry, the situation is more complicated, as there is interaction [20]. However the dimension of the variables still correspond and so we continue to assume $\rho \leftrightarrow$ Q.

Next, in our case, we are interested in Q packet in $T_{dura}$, where Q=i*$T_{dura}$. Thus we use i instead of Q or:

$$\rho \leftrightarrow i. \tag{99}$$

It is also noted that both $\rho$ and i have the following property:
a) Order parameter and fluctuates
b) $\rho \rightarrow$ number of particles; but here number of net charges $\rightarrow$ normalized by time becomes current i (consistency check: charges (particles) correspond to particles in vdw gas, so charge particles $\rightarrow$ N $\rightarrow$ i) [21]

T=?

In vdw gas, T is used to represent fluctuation/randomness (remember by definition T= $\partial E/\partial S$, with S=entropy or randomness, also from kT, is the thermal energy/randomness, we note T represent fluctuation in energy of the state $E_v$, due to thermal energy kT). Also T is the integrating factor of the heat transferred from the heat bath/reservoir [35]. Here with thermal noise there is fluctuation in energy of the charge. Thus T (via kT) is also used to represent fluctuation in energy of the ensemble of charge on top plate when M2 just hits off. Again T is the integrating factor of the heat transferred from the heat bath/reservoir, which is the resistor R. Note T=temperature (not period). Thus

---

[19] Thus pressure $\leftrightarrow$ voltage, volume $\leftrightarrow$ C or capacitor.

[20] Similarly with the analogy, because now it is compared with container of vdw gas, again there is interaction.

[21] What corresponds to N? The ensemble of species/packet of charge Q, in duration T, becomes, from i=dq/dt, Q=∫idt $\propto$ i(or 2$I_o$=dq/dt, Q=∫2$I_o$dt $\propto$ 2$I_o T_{dura}$, then we have 2$I_o \leftrightarrow$ N

$$T \leftrightarrow T \tag{100}$$

From (98), (99), (100) partition function for packet Q, is:

$$\Xi\ (\mu= E_{ft} = E_{fb}, \rho,\ T) \leftrightarrow \Xi\ (\mu= E_{ft} = E_{fb}, i, T) \tag{101}$$

We can follow Mayer cluster expansion [35] and find equation of state [22], and then find fluctuation. However, after we get fluctuation from partition function, we need to do scaling (follow [53]) by multiplying the number of modes (follow [53]) to convert fluctutation in N to i. Also since we measure not fluctuation in i but $i^2/\Delta f$, then the scaling is

$$kT\ \leftrightarrow 4kT/R \tag{102}$$

Specifically using charge on capacitor C in Figure 1, the resemblance to vdw gas/Ising neighboring spin can be seen.

The charge, having thermal energy kT, when flowing as current through R, dissipates Joule heat, pretty much like vdw gas dissipates heat in the reservoir. Following [53], the power spectral density (psd) of this heat is given as 4kT/R. Thus the noise term in vdw gas, β or kT, has its counterpart here, except it is scaled by R and is 4kT/R. [23]

REFERENCES

[1]   Y. Cao, P. Leroux, W. Cock, M. Steyaert, "A 1.7mW 11b 1-1-1 MASH Time-to_Digital Converter", Digest ISSCC, pp. 480-481, San Francisco, Feb 2011.
[2]   A. Buonomo, Analysis of Source-Coupled Multivibrators. IEEE Transactions on Circuits and Systems, pp. 1197 – 1200. June 2006
[3]   Behzad Razavi, "**A** Study of Phase Noise in CMOS Oscillators", IEEE Journal of Solid State Circuits, Vol. 31, No.32, pp. 331-343, ,March, 1996.
[4]   A. Hajimiri, T. H. Lee "A General Theory of Phase Noise in Electrical Oscillators", IEEE Journal of Solid State Circuits, Vol. 33, No. 2, pp. 179-194, Feb 1998.
[5]   R. Navid. T. Lee, Dutton, "Minimum Achievable Phase Noise of RC oscillators," IEEE Journal of Solid-State Circuits, vol. 40, no. 3, pp. 630-637, March 2005
[6]   R. Navid, T. H. Lee and R. W. Dutton, "Circuit-Based Characterization of Device Noise Using Phase Noise Data," *IEEE Transactions on Circuits and Systems* I, Vol. 57, pp. 1265-1272, June 2010.
[7]   A. Abidi, R. Meyer, "Noise in Relaxation Oscillators", IEEE Journal of Solid State Circuits, Vol. 18, No. 6,   pp. 794-802, December 1983.
[8]   S. Shankar Sastry, "The Effects of Small Noise on Implicitly Defined Nonlinear Dynamical Systems", IEEE Transactions on Circuits and Systems, Vol. 30, No. 9, pp. 651-663, September 1983
[9]   S. Sastry, O. Hijab, "Bifurcation in the presence of small noise", Systems and Control Letters, Vol. 1, No. 3, PP. 159-167, November 1981
[10]  B. Leung, "A Switching Based Phase Noise Model for CMOS Ring Oscillators Based on Multiple Thresholds Crossing", IEEE Trans on Circuits and System I, pp.2858-2869, Nov 2010
[11]  Sedigheh Hashemi and Behzad Razavi, "Analysis of Metastability in Pipelined ADCs", IEEE Journal of Solid-State Circuits, vol. 49, no. 5, pp. 1198-1209, May, 2014
[12]  P. Figueiredo, "Comparator Metastability in the Presence of Noise", IEEE Trans on Circuits and System I, pp.1286-1299, May 2013
[13]  Pierluigi Nuzzo, Fernando De Bernardinis, Pierangelo Terreni, and Geert Van der Plas, "Noise Analysis of Regenerative Comparators for Reconfigurable ADC Architectures", IEEE Transactions on Circuits and Systems I, pp. 1441-1454, July 2008
[14]  S. K. Mathew, S. Srinivasan, and M. A. Anders, "2.4 Gbps, 7 mW all digital PVT-variation tolerant true random number generator for 45 nm CMOS high-performance microprocessors," IEEE Journal of Solid-State Circuits, vol. 47, no. 11, pp. 2807–2821, November 2012.
[15]  N. G. Vampen, "Stochastic Processes in Physics and Chemistry", 3rd edition, North-Holland/Elsevier, 2007
[16]  H. Callen, "Thermodynamics, Wiley, 1960
[17]  S. Strogatz, "Nonlinear dynamics and chaos", Addison-Wesley, 1994
[18]  P. M. Morse, H. Feshbach, Methods of Theoretical Physics I, McGraw Hill, New York, 1953.
[19]  M. Freidlin, A. Wentzell, "Random Perturbations of Dynamical Systems", Springer Verlag, 1984
[20]  L. Landau, E. Lifshitz, "Statistical Physics", volume 5 of course of theoretical physics, Pergamon Press, Addison-Wesley, 1958.
[21]  B. Leung, "Noise/Jump Phenomenon of Relaxation Oscillators Based on Phase Change Using Path Integral/Lagrangian Formulation in Quantum Mechanics", Midwest Symposium on Circuits and Systems, Boise, Idaho, pp. 246-249, Aug 2012
[22]  B. Leung, "Noise Model of Relaxation Oscillators Using Rayleigh Dissipation Function and Lagrangian with Friction", IEEE International Conference on Circuits and Systems, Kula Lumper, Malaysia pp.9-12, September, 2013
[23]  Prof. Leggett, University of Illinois, Urbana, private communication
[24]  H. Goldstein, "Classical Mechanics", Addison-Wesley, 1950
[25]  "Lagrangians and Hamiltonians with friction", Journal of Physics: Conference Series vol 237, pg. 012-021, 2010
[26]  Prof. Mathis, Institute of Theoretical Electrical Engineering, Leibniz University  of Hannover), private communication

---

[22] Of course, equation will be like vdw equation of state.  On the other hand, if we had derived equation of state using  KCL, KVL i.e. conservation of charge/energy, with interaction in vdw gas now represented by nonlinear I-V characteristics of M1, M2, the 2 approaches should give the same equation. Now in partition function/Mayer decomposition approach, the linear coefficient has dependency on T, temperature.  With KCL, KVL, T does not show up explicitly in the linear coefficient, which is $g_mR$. We need to add normalization in  section IV.C 1)  to make it appears like $T/T_c$. This is because vdw's equation of state starts from details with noise/thermal energy (partition function) and then take the mean, while KCL, KVL assume deterministic (no noise, no detail) to begin with. Thus the vdw coefficient has more detail (like T in it). In a sense it is like ac/small signal equation, but with coefficient (e.g. $g_m$) depends on  dc biasing condition ($I_D$); here vdw equation (taking mean/average), has coefficient depending on thermal energy (kT).

[23] We have now assumed the noise term is dominated by R. To be more accurate, we can use (29), i.e. $\lambda=4kT/R+4kT\gamma g_m$. Finally if we want to include all parameters we have $\lambda=4kT/R+4kT/R+ 4kT\gamma g_{m1}+4kT\gamma g_{m2}$. Since near jump point, only 1 of the transistor (assume M2) is on, and using long channel approximation for γ, we have noise coming from M2 together with its associated resistor R only. Thus the noise term should be $4kT(R+1/(2/3 \times g_m))$.

[27] L. Weiss, W. Mathis, "A Thermodynamic Approach to Noise in Non-Linear Network", International Journal of Circuit Theory and Application, pg. 147-165, 1998.
[28] A. Papoulis and S. Pillai, *Probability, Random Variables and Stochastic Processes*. Upper Saddle River, NJ: Prentice Hall, 2002.
[29] H. Kleinert, "Path Integrals in Quantum Mechanics, Statistics and Polymer Physics", World Scientific, 1995
[30] http://en.wikipedia.org/wiki/Wick_rotation
[31] C. E. Smith, "Lagrangians and Hamiltonians with friction", Journal of Physics: Conference Series vo237, , pg. 012021, 2010
[32] B. Leung, " Novel dissipative Lagrange-Hamilton formalism for LC/van der pol oscillator with new implication on phase noise dependency on quality factor", Midwest Symposium on Circuits and Systems, Texas, Aug 2014, pg. 507-510
[33] V. Arnold, "Ordinary Differential Equations", 1973, MIT Press
[34] D. Chandler, "Introduction to Modern Statistical Mechanics", Oxford University Press, 1987
[35] R.K. Pathria, "Statistical Mechanics", Pergamon Press, 1972
[36] A. S. Sedra and K. C. Smith, Microelectronic Circuits. Oxford University Press, 2010.
[37] J. Negele, H. Orland, "Quantum Many-Particle Systems", Addison Wesley, 1987
[38] R. Shankar, "Principles of Quantum Mechanics", Kluwer Academic Publisher, 1994.
[39] Feynman and Hibbs, "Quantum Mechanics and Path Integrals: Emended Edition", McGraw-Hill, New York, 1965
[40] Feynman and Vernon, "The theory of a general quantum system interacting with a linear dissipative system", Annals of Physics, Volume 24, October 1963, Pages 118–173, Elsevier
[41] Feynman, "Statistical Mechanics", A set of Lectures", W. Benjamin, Inc., 1972
[42] Quantum dissipative systems, II: The influence-functional approach
[43] Phys504 – Statistical Physics – Fall 2006 Lecture #24 Professor Anthony J. Leggett, Department of Physics, UIUC
[44] Caldeira and Leggett, "Path integral approach to quantum Brownian motion Physica A: Statistical Mechanics and its Applications Volume 121, Issue 3, September 1983, Pages 587–616 North-Hollund Publishing Co.
[45] Diosi, "Caldeira and Leggett Master Equation and Medium Temperature", Physics A, 199, 1993, 517-526, North Holland.
[46] M. Razavy, "Classical and quantum dissipative systems", Imperial College Press, "2005.
[47] Reza Navid, Thomas H. Lee, Robert W. Dutton, "Circuit-Based Characterization of Device Noise Using Phase Noise Data" IEEE Transactions on Circuits and Systems, pp. 1265-1272. June 2010
[48] Schulman, "Techniques and applications of path integration", Dover Publication, New York, 2005
[49] Kittel, "Introduction to solid state physics", 4th edition, Wiley and Sons, 1971
[50] Mckelvey, "Solid state and semiconductor physics", Robert Krieger Publishing Company, 1982
[51] Muller and Kamins, "Device electronics for integrated circuits", Wiley and Sons, 1977
[52] S. Datta, "Electronic transport in mescocopic systems", Cambridge, 1995.
[53] Middleton, "Introduction to statistical communication theory", McGraw Hill, 1960
[54] Wikipedia: GCE, grand canonical ensemble.
[55] J. Blakemore, "Semiconductor Statistics", Pergamon, 1962
[56] J. Stoker, "Nonlinear Vibrations", Interscience Publishers, 1966